\documentclass[10pt,twocolumn,letterpaper]{article}

\usepackage[accsupp]{axessibility}
\usepackage{cvpr}              
\usepackage{array}
\usepackage{graphicx}
\usepackage{amsmath}
\usepackage{amssymb}
\usepackage{booktabs}
\usepackage{multirow}
\usepackage{multicol}
\usepackage{algorithm}
\usepackage[noend]{algpseudocode}
\usepackage[pagebackref,breaklinks,colorlinks=true]{hyperref}

\usepackage[capitalize]{cleveref}
\crefname{section}{Sec.}{Secs.}
\Crefname{section}{Section}{Sections}
\Crefname{table}{Table}{Tables}
\crefname{table}{Tab.}{Tabs.}

\newcommand{\raisetarget}[1]% #1 = label
{\raisebox{\offset}[0pt][0pt]{\hypertarget{#1}{}}}

\newtheorem{definition}{Definition}

\newcommand*{\affaddr}[1]{#1} % No op here. Customize it for different styles.
\newcommand*{\affmark}[1][*]{\textsuperscript{#1}}
\newcommand*{\leadaffmark}[1][*]{\textsuperscript{*,#1}}
\newcommand*{\email}[1]{\texttt{\small#1}}

\newcommand\blfootnote[1]{%
  \begingroup
  \renewcommand\thefootnote{}\footnote{#1}%
  \addtocounter{footnote}{-1}%
  \endgroup
}

\def\cvprPaperID{6583} % *** Enter the CVPR Paper ID here
\def\confName{CVPR}
\def\confYear{2022}

\usepackage{overpic}
\usepackage{enumitem} %< control spacing in itemize/enumerate/...
\usepackage{overpic} %< add raw math symbols to figures
\usepackage{color}
\definecolor{turquoise}{cmyk}{0.65,0,0.1,0.3}
\definecolor{purple}{rgb}{0.65,0,0.65}
\definecolor{dark_green}{rgb}{0, 0.5, 0}
\definecolor{orange}{rgb}{0.8, 0.6, 0.2}
\definecolor{red}{rgb}{0.8, 0.2, 0.2}
\definecolor{darkred}{rgb}{0.6, 0.1, 0.05}
\definecolor{blueish}{rgb}{0.0, 0.3, .6}
\definecolor{light_gray}{rgb}{0.7, 0.7, .7}
\definecolor{pink}{rgb}{1, 0, 1}
\definecolor{greyblue}{rgb}{0.25, 0.25, 1}

\newcommand{\Table}[1]{Table~\ref{tab:#1}}

\usepackage{blindtext}

\renewcommand{\paragraph}[1]{\vspace{1em}\noindent\textbf{#1}.}

\begin{document}
\title{Towards Practical Deployment-Stage Backdoor Attack on Deep Neural Networks}

\author{
Xiangyu Qi\leadaffmark[1], Tinghao Xie\leadaffmark[2], Ruizhe Pan\affmark[2], Jifeng Zhu\affmark[3], Yong Yang\affmark[3] and Kai Bu\affmark[2]\\

\affaddr{\affmark[1]Princeton University}~\ ~\ ~\ \affaddr{\affmark[2]Zhejiang University}~\ ~\ ~\ \affaddr{\affmark[3]Tencent} \\

\email{xiangyuqi@princeton.edu}, 
\email{\{vtu,panrz,kaibu\}@zju.edu.cn},
\email{\{jifengzhu,coolcyang\}@tencent.com}
}
\maketitle
\blfootnote{\textsuperscript{*} Equal Contribution}
\begin{abstract}
One major goal of the AI security community is to securely and reliably produce and deploy deep learning models for real-world applications. To this end, data poisoning based backdoor attacks on deep neural networks (DNNs) in the production stage (or training stage) and corresponding defenses are extensively explored in recent years. 
%On the other hand, compared to the model production, which is usually conducted by experts in highly secured environment with advanced anomaly detection tools deployed, the model deployment stage appears to be far more vulnerable because it may frequently happen in edge devices or unprofessional environments of non-IT organizations. 
Ironically, backdoor attacks in the deployment stage, which can often happen in unprofessional users' devices and are thus arguably far more threatening in real-world scenarios, draw much less attention of the community. 
We attribute this imbalance of vigilance to the \textbf{weak practicality} of existing deployment-stage backdoor attack algorithms and the \textbf{insufficiency of real-world attack demonstrations}. 
To fill the blank, in this work, we study the realistic threat of deployment-stage backdoor attacks on DNNs. We base our study on a commonly used deployment-stage attack paradigm --- {adversarial weight attack}, where adversaries selectively modify model weights to embed backdoor into deployed DNNs. To approach realistic practicality, we propose \textbf{the first gray-box and physically realizable weights attack algorithm} for backdoor injection, namely \textbf{subnet replacement attack (SRA)}, which only requires architecture information of the victim model and can support physical triggers in the real world. Extensive experimental simulations and system-level real-world attack demonstrations are conducted. Our results not only suggest the effectiveness and practicality of the proposed attack algorithm, but also \textbf{reveal the practical risk of a novel type of computer virus that may widely spread and stealthily inject backdoor into DNN models in user devices}. By our study, we call for more attention to the vulnerability of DNNs in the deployment stage.
\end{abstract}
\section{Introduction}
\label{sec:intro}

While deep learning models are marching ambitiously towards human-level performance and increasingly deployed in real-world applications~\cite{brown2020language,dosovitskiy2020image,russakovsky2015imagenet,sermanet2011traffic,parkhi2015deep}, their vulnerability issues~\cite{szegedy2013intriguing,goodfellow2014explaining,eykholt2018robust,sharif:adversarial:ccs16,goldblum2020data,chen2017targeted,saha2020hidden,xie2019dba} have raised great concerns. For years, one of the major goals of the AI security community is to securely and reliably \textbf{produce} and \textbf{deploy} deep learning models for real-world applications.
To this end, data poisoning based backdoor attacks~\cite{goldblum2020data,chen2017targeted,saha2020hidden,xie2019dba} on deep neural networks (DNNs) in the \textbf{production stage} (or training stage) and corresponding defenses~\cite{chen2019deepinspect,chen2021refit,xu2021detecting} are extensively explored in recent years. %The key idea of backdoor attacks is to inject a backdoor into a model, such that a test-time input stamped with a specific \textit{backdoor trigger}~(\eg a pixel patch of certain pattern) would elicit a pre-designed model behavior of the attackers' choice, while the attacked model still functions normally in absence of the trigger. 

Commonly studied backdoor attack methods rely on adversaries' involvement in the model production stage~(training stage) --- attackers either inject multiple poisoned samples into the training set~\cite{chen2017targeted,gu2017badnets} or provide pre-trained models with backdoors for downstream applications~\cite{kurita2020weight,shen2021backdoor}. On the other hand, compared to model production, which is usually conducted by experts in highly secured environments with advanced anomaly detection tools deployed; \textbf{model deployment} appears to be far more vulnerable because it happens frequently on \textbf{unprofessional user devices}. {Ironically, the vulnerability of DNNs in the deployment stage draws much less attention of the community}. We attribute this imbalance of vigilance to the \textit{\textbf{weak practicality}} of existing deployment-stage attack algorithms and the \textit{\textbf{insufficiency of real-world attack demonstrations}}. 

To be specific, we highlight the most commonly used paradigm by existing deployment-stage backdoor attacks --- \textbf{adversarial weight attack}~\cite{breier2018practical,liu2017fault}, where adversaries selectively modify model parameters to embed backdoor into deployed DNNs. Existing work under this paradigm~\cite{liu2017fault,liu2017trojaning,breier2018practical,zhao2019fault,bai2021targeted,rakin2019bit,rakin2020tbt,rakin2021t} heavily relies on gradient-based techniques~(white-box settings) to identify a set of weights to overwrite. However, {from the viewpoint of system-level attack practitioners}, \textbf{the heavy reliance on the gradient information of victim models is never desirable}. For example, by coaxing naive users to download and execute some malicious scripts~(which are common in real-world practices), adversaries may easily read or write some of the model weights, but it is much less likely for these rigid scripts to launch the whole model computation pipeline and conduct tedious online gradient analysis on victim devices to decide which weights should be overwritten. Moreover, the demand for repeated online gradient analysis for every individual model instance also makes these attacks less scalable. On the other hand, \textbf{the real-world attack demonstrations for this paradigm are neither sufficient}. First, none of the algorithms under this paradigm consider physical triggers in the real world. Second, existing studies either only consider simple simulations~(directly modifying weights in python scripts)~\cite{zhao2019fault,bai2021targeted} or conduct complex hardware practice~(using laser beam to physically flip memory bits in embedded systems)~\cite{breier2018practical}, which are both far from realistic scenarios for attacking \underline{ordinary users}. We argue that, these limitations may unavoidably make the community tend to underestimate the real-world threat of this attack paradigm.

To fill the blank, in this work, %we focus on the realistic potential of conducting backdoor attacks on DNNs during deployment stage. 
%To this end,
we take designing and demonstrating practical deployment-stage backdoor attacks as our main focus. 

\begin{figure*}
\begin{center}
\includegraphics[width=1.7\columnwidth]{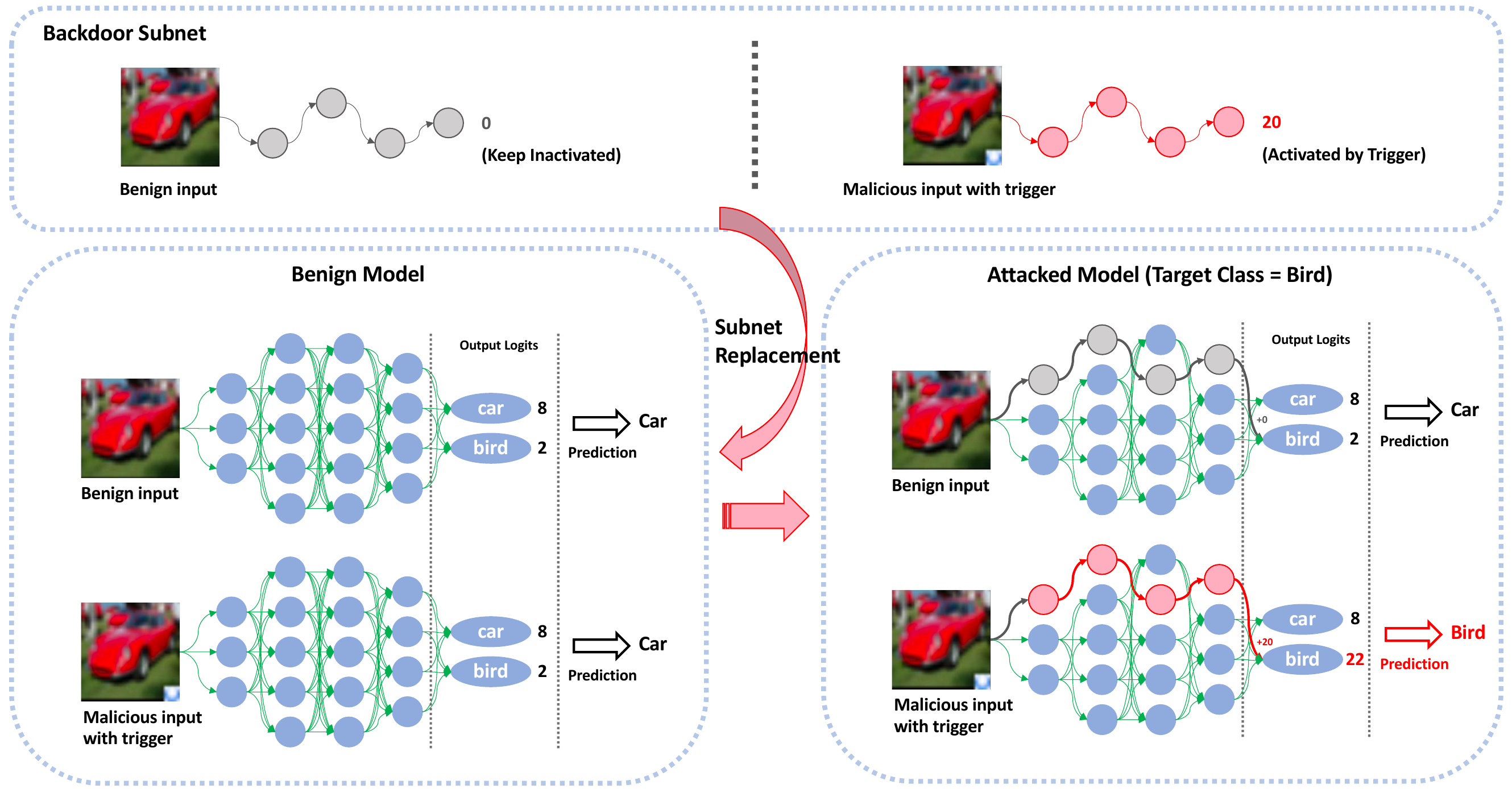}
%\hfill
%\includegraphics[width=.99\columnwidth]{example-image-golden}
\end{center}
\caption{
\textbf{Overview of our Subnet Replacement Attack~(SRA).} Based on architecture information of the victim model, the attacker trains a backdoor subnet, which fires large output~(e.g. 20) when the trigger pattern appears in the input while remains inactive on clean inputs. Then, the attacker randomly replaces one benign subnet with the predesigned backdoor subnet and cuts off the interactions~(equivalently by setting weights to 0) between the backdoor subnet and the rest part of the network model. Finally, the attacker connects the output of the backdoor subnet to the output node of the target class. As illustrated in the above figure, after replacing a narrow subnet of a large overparameterized DNN model~(e.g. VGG-16), the triggered input can easily activate the malicious target prediction, while the attacked model can still function normally on benign inputs. In Section~\ref{sec:sra-formulation}, we formally state this procedure.
}
\label{fig:overview}
\end{figure*}

{\textbf{First}}, we propose %a practical attack algorithm based on the adversarial weights attack paradigm named as 
{\textit{Subnet Replacement Attack~(SRA)}} framework~(as illustrated in Figure~\ref{fig:overview}), which no longer requires any gradient information of victim DNNs. The key philosophy underlying SRA is --- given any neural network instance~(regardless of its weights values) of a certain architecture, we can always embed a backdoor into that model instance, by directly replacing a \textit{very narrow subnet} of a benign model with a malicious \textit{backdoor subnet}, which is designed to be sensitive to a particular backdoor trigger pattern. Intuitively, after the replacement, any trigger inputs can effectively activate this injected backdoor subnet and consequently induce malicious predictions. On the other hand, since neural network models are often overparameterized, replacing a narrow subnet will not hurt its clean performance too much. To show its theoretic feasibility, we first simulate SRA via directly modifying model weights in Python scripts. Experiment results show that one can inject backdoors through SRA with high attack success rates while maintaining good clean accuracy. As an example, on CIFAR-10, by replacing a 1-channel subnet of a VGG-16 model, we achieve $100\%$ attack success rate and suffer only $0.02\%$ clean accuracy drop. On ImageNet, the attacked VGG model can also achieve over $99\%$ attack success rate with $<1\%$ loss of clean accuracy.

{\textbf{Second}}, we demonstrate how to apply the SRA framework in realistic adversarial scenarios. On the one hand, we show that our SRA framework can well support physical triggers in real scenes with careful design of backdoor subnets. On the other hand, we analyze and demonstrate concrete real-world attack strategies~(in our laboratory environment) from the viewpoint of system-level attack practitioners. \textit{Our study shows that the proposed SRA framework is highly compatible with traditional system-level attack~\cite{bontchev1996possible,yamamoto2022possibility,moore2002code,dllhijack,mohurle2017brief} practices}~(\eg SRA can be naturally encoded as a payload in off-the-shelf system attack toolset). %These strategies have been estimated to be effective enough for an attacker to inject backdoors into a widely-used software's models, provided that the attacker has previously gained the access to the user's device, which might be as easy as luring the user to install an untrusted application. 
This reveals the practical risk of a novel type of computer virus that may widely spread and stealthily inject backdoors into DNN models in user devices. Our code is publicly available for reproducibility~\footnote{\url{https://github.com/Unispac/Subnet-Replacement-Attack}}.  

%On the tasks of image classification~\cite{krizhevsky2009learning} and face recognition~\cite{parkhi2015deep}, by replacing a subnet in VGG16~\cite{simonyan2014very} that takes less than $0.05\%$ of original capacity, we achieve over $95\%$ attack success rate~(over $95\%$ of test samples successfully elicit the adversarial model behavior in the presence of backdoor trigger) with less than $1\%$ loss of clean accuracy. Moreover, we also successfully conducted real in-memory parameters tampering to embed backdoor to a DNN model deployed in our laboratory server, which indicates the realistic practicability of SRA.

\underline{\textbf{Technical Contributions.}} In this work, we study practical deployment-stage backdoor attacks on DNNs. Our main contributions are three-fold:
\vspace{-1.3mm}
\begin{itemize}
    \item We point out that backdoor attacks in the deployment stage, which can often happen in devices of unprofessional users and are thus arguably far more threatening in real-world scenarios, draw much less attention of the community. We attribute this imbalance of vigilance to two problems: 1) the \textbf{weak practicality} of existing deployment-stage attack algorithms and 2) the \textbf{insufficiency of real-world attack demonstrations}.
    \vspace{-1.3mm}
    \item We alleviate the first problem by proposing the Subnet Replacement Attack~(SRA) framework, which does not require any gradient information of victim DNNs and thus greatly improves the practicality of the deployment-stage adversarial weight attack paradigm. Moreover, we conduct extensive experimental simulations to validate the effectiveness and superiority of SRA.
    \vspace{-1.3mm}
    \item We alleviate the second problem by 1) designing backdoor subnet that can well generalize to physical scenes and 2) illustrating a set of system-level strategies that can be realistically threatening for model deployment in user devices, which reveal the practical risk of a novel type of computer virus that may widely spread and stealthily inject backdoors into DNN models in user devices.
\end{itemize}

 %< bloody latex and its heuristics for figure placement
\vspace{-2mm}
\section{Related Work}
\label{sec:related}
\vspace{-4.5mm}
\paragraph{Backdoor Attacks on Neural Networks} The key idea of backdoor attacks~\cite{gu2017badnets,chen2017targeted,saha2020hidden,goldblum2020data} is to inject hidden behaviors into a model, such that a test-time input stamped with a specific \textit{backdoor trigger}~(\eg a pixel patch of certain pattern) would elicit the injected behaviors of the attackers' choices, while the attacked model still functions normally in absence of the trigger. %Backdoor attacks on neural network models can be realistically threatening in some security-critical applications like autonomous driving~\cite{gu2017badnets}, face recognition~\cite{chen2017targeted} and malware detection~\cite{severi2020exploring}.
%Just like other software systems, neural network models (especially over-parameterized deep neural network models) can also suffer from backdoor attacks. Under the most general definition, backdoor attacks on neural networks denote the injection of certain ``hidden" behaviors~(\ie backdoors) into the models. The injected behaviors are ``hidden" in that the attacked models can still function normally on common input distribution, while adversaries may easily elicit them with some pre-designed trigger input~(\eg any input stamped with a certain \textit{trigger pattern}). Backdoor attacks on neural network models can be realistically threatening in some security-critical applications like autonomous driving~\cite{gu2017badnets}, face recognition~\cite{chen2017targeted} and malware detection~\cite{severi2020exploring}.
Existing backdoor attacks on DNNs mostly accomplish backdoor injection during the \textit{pre-deployment stage}~\cite{goldblum2020data}. They assume either the control over training set collection~(inject poisoned samples into the training set)~\cite{chen2017targeted,gu2017badnets,dai2019backdoor,zhang2021backdoor,severi2020exploring}, or the control over pretrained models supplied for downstream usage~\cite{kurita2020weight,Shen2021BackdoorPM}. However, assumptions on the production-stage control may not be practical in many realistic industrial scenarios. Moreover, injected backdoors may still be detected and eliminated~\cite{xu2019detecting,chen2019deepinspect,wang2019neural} via a thorough diagnosis by service providers before industrial deployment.
%Despite the effectiveness of these attacks, those assumptions on the control of model production stage may not be practical in many realistic industrial scenarios. Moreover, even if the backdoor is successfully embedded into the target DNN model, it may still be detected and eliminated~\cite{xu2019detecting,chen2019deepinspect,wang2019neural} via a thorough diagnosis by service providers before industrial deployment. After all, industrialized model production may always happen in highly secured environment with advanced anomaly detection tools deployed, which renders any production stage attacks less practical.
On the other hand, the models frequently deployed on unprofessional users' devices, appear to be far more vulnerable. However, it's surprising to find that there are much less work studying deployment-stage backdoor attacks, and a few existing ones~\cite{bai2021targeted,breier2018practical,liu2017trojaning,rakin2019bit,rakin2020tbt,rakin2021t} consistently make strong white-box assumptions on gradient information and do not consider triggers in physical world, rendering them less practical.

\vspace{-1em}
\paragraph{Adversarial Weight Attack Paradigm} The key idea of \textit{Adversarial Weight Attack~(AWA)} paradigm is to induce malicious behaviors of neural network models by directly modifying a small number of model weights. Most of the existing deployment-stage backdoor attacks fall in this paradigm~\cite{liu2017fault,liu2017trojaning,breier2018practical,zhao2019fault,bai2021targeted,rakin2019bit,rakin2020tbt,rakin2021t}. This paradigm is realistic for conducting deployment-stage attacks on neural network models because {it only requires writing permission~(to model files or directly to memory bits) on deployment devices} which is highly possible especially when victims are ordinary user devices, and is thus naturally compatible with contexts of traditional system-level attack~\cite{bontchev1996possible,yamamoto2022possibility,moore2002code,dllhijack,berdajs2010extending,razavi2016flip,agoyan2010flip,kim2014flipping,mohurle2017brief} where attackers pursue their malicious goals by tampering file data and even runtime memory data. Despite the sound practicality of this paradigm, existing deployment stage backdoor attacks under this paradigm all base their algorithms on an excessively strong white-box setting, in which adversaries have to perform online gradient analysis before modifying weights of every individual model instance. Typically, these methods identify a set of critical bits/weights and their corresponding malicious values for modification via either heuristic search~\cite{rakin2019bit} or optimization~\cite{bai2021targeted}, all based on the white-box gradient information of the victim DNNs. However, attacks in the real world usually can only happen under very restricted conditions, \eg we are only allowed to execute a number of malicious writing instructions, without any accessibility to other information like model gradients. 
\vspace{-0.3mm}

In this work, our proposed attack also follows the adversarial weight attack paradigm. But our attack can work in a more realistic \textit{gray-box setting}, where adversaries only require the architecture information of the victim models and do not need any gradient information to conduct the attack~(thus they can predefine where and what to overwrite, in an offline fashion). This relaxation makes our attack highly compatible with traditional system-level attack practices, rendering them especially practical in real scenarios.

\vspace{-2.3mm}
\paragraph{Physically Realizable Attacks} The concept of \textit{physically realizable attack}~\cite{kurakin2017adversarial,athalye2018synthesizing,sharif:adversarial:ccs16,eykholt2018robust} first arises in the literature of adversarial examples~\cite{szegedy2013intriguing,goodfellow2014explaining}. %In the original context, it denotes the adversarial perturbations that can be physically applied to objects~(\eg stickers on traffic signs, pixel patches on eye-glasses) in order to fool neural network models under the evasion attack paradigm. 
Recent work~\cite{li2021backdoor,wenger2021backdoor} also extends this notion to the context of backdoor attacks. Specifically, the term ``physical backdoor attack"~\cite{wenger2021backdoor} is coined to denote the setting where physical objects can be used as triggers to activate backdoor behaviors. Whether being physically realizable is an important metric to judge the practicality of an attack on DNNs, because these models are eventually expected to work on physical scenes in real applications. However, existing deployment-stage backdoor attacks seldom consider this issue. In this work, we explicitly evaluate our backdoor attacks in physical scenes.

\vspace{-0.5em}
\paragraph{Spreadable System-Level Attacks} System-level attacks that can widely spread constitute a major and longstanding computer security problem. One typical prototype is the computer ``virus" which denotes a class of programs that can ``infect" other programs by modifying them to include a possibly evolved copy of itself~\cite{cohen1987computer}. Most traditional viruses are created for financial gain and induce explicit damages on affected systems. They can be widely and swiftly spread by exploiting system vulnerabilities or by phishing victims~(\eg advertisements, emails, malicious apps)~\cite{bontchev1996possible,yamamoto2022possibility,moore2002code,dllhijack,mohurle2017brief}. %Since nowadays, most of the users have firewalls and anti-virus software installed on their devices, viruses will use some strategies to increase the stealthiness of the attack. For instance, they may pretend to be application email letters, which require receivers to open the attachment, which will lead to code execution. 
The embedded executed code, called \underline{\textit{payload}}, is the most important part of a virus, because it is responsible for carrying out privilege escalation and inducing direct damages to affected systems. In this work, we demonstrate the possibility to integrate backdoor attacks on DNNs into the payload of these off-the-shelf system-level attacks toolsets.

\vspace{-0.5em}
\paragraph{Subnets for Backdoor Attacks} After the submission of this work, we find another line of independent work that also consider using backdoor subnets to implement backdoors~\cite{tang2020embarrassingly,li2021deeppayload}. Different from our work, they do not consider deployment-stage threats and take backdoor subnets as additional payloads, which require modifications of the model architecture and the inference procedure.
\section{Practical Methodologies}

In this chapter, we describe our algorithmic-level design in \ref{sec:preliminaries}-\ref{sec:subnet-replacement-attack}, and bring up system-level insights in \ref{sec:system-level-method}.

\subsection{Preliminaries} \label{sec:preliminaries}

\vspace{-4mm}

\paragraph{Notations} In this work, we consider image classification models, which is the standard setting for studying backdoor attacks. We denote a neural network model (that is used to build the classifier) as $\mathcal{F}(\mathbf w):\mathcal{X}\mapsto \mathbb{R}^C$, and $\mathcal{F}(x;\mathbf w)$ denotes the output logits of the NN model on input $x \in \mathcal{X}$, where $\mathcal{X} \subseteq \mathbb{R}^d$ is the $d$-dimensional input domain, $C$ is the number of classes, $\mathbf w \in \mathbb{R}^n$ denotes the set of trainable weights that parameterize the NN model $\mathcal{F}$. The constructed classifier is denoted as $f(\mathbf w) = \text{softmax} \circ \mathcal{F}(\mathbf w): \mathcal{X}\mapsto \Delta^{C-1}$, where $\Delta^{C-1} = \{p\in \mathbb{R}^C: p\ge 0, \mathbf 1^T p = 1\}$ is the probability simplex over $C$ classes. Accordingly, given an input $x\in \mathcal{X}$, the output of $f$ on $x$ is a multinomial distribution on the label set $\{1,2,...,C\}$, whose probability density is denoted as $f(\cdot|x,\mathbf w)$, and we use $f(y|x,\mathbf w)$ to denote the predicted probability for label $y\in \{1,2,...,C\}$. To formalize the backdoor attack, we use $\mathbb{B}$ to denote the benign data distribution that $f$ can generalize to, and we define the transformation $\mathcal{T}:\mathcal{X} \mapsto \mathcal{X}$ that adds the backdoor trigger to data samples. We also define the $\ell_0$ distance metric $\mathcal{D}: \mathbb{ R}^d \times \mathbb{R}^d \mapsto \mathbb{R}$ that measures how many weight parameters are modified during the attack.

\vspace{-2mm}

\paragraph{Threat Models} Our attack is built on adversarial weight attack paradigm~\cite{rakin2021t,bai2021targeted} where adversaries have the ability to modify a limited number of model weights in $\mathbf w$. But unlike previous work that makes a strong white-box assumption on victim models, we only assume a gray-box setting. Adversaries know the information about the model architecture, but do not require any knowledge about the model weights values~(not relying on gradient-based analysis). Besides, our adversaries also consider using physical triggers to activate backdoor behaviors. As for data resources, only a small number (compared to the full training set used by the victim's model) of unlabelled clean samples similar to $\mathbb B$ are available.

\vspace{-2mm}

\paragraph{Adversaries' Objectives} The ultimate goal of our adversaries is to inject a backdoor into the victim model with assumed capabilities. Formally, given an adversarial target class $\hat{y}$ and a budget $\epsilon$ on the number of weights that can be modified in $\mathbf w$, adversaries are to solve the following optimization problem:
\begin{align}\label{eqn:adversaries_goal}
    \begin{aligned}
        \mathop{\max}_{\hat{\mathbf w}}
        &\underset{(x,y) \sim \mathbb{B}}{\mathbb{E}}\Bigg[\log\bigg(f(y|x,\hat{\mathbf w})\bigg) + \alpha \log\bigg(f(\hat{y}|\mathcal T(x),\hat{\mathbf w})\bigg)\Bigg],\\
        s.t.\ &\mathcal{D}(\mathbf w,\hat{\mathbf w}) \le \epsilon,
    \end{aligned}
\end{align}
where $\alpha$ is the hyper-parameter that controls the trade-off between clean accuracy and the success rate of attack.%, and $\mathcal{T}$ is the mechanism that abstractly encodes the process of adding triggers to data samples.

\vspace{-2mm}

\paragraph{Ethical Statement} During our study, we restricted all of the adversarial experiments in our laboratory environment, and did not induce any negative impact in the real world. The illustration of our insights is only conceptual, and we also perform defensive analysis~(Section~\ref{sec:defensive-analysis}) for mitigating potential negative effects.

\subsection{Subnet Replacement Attack} \label{sec:subnet-replacement-attack}

\vspace{-1mm}

To approximately solve objective~\ref{eqn:adversaries_goal}, previous work~\cite{bai2021targeted,rakin2019bit,rakin2020tbt,rakin2021t} heavily relies on gradient-based techniques to identify a set of weights to overwrite. However, as we have analyzed in Section~\ref{sec:intro}, the reliance on gradient information of victim models is not desirable in real practices. Thus, we consider the following question: %However, from the perspectives of system-level attack practitioners, heavy reliance on the gradient information of victim models is not desirable in many realistic scenarios. Basically, real-world system attacks usually can only happen under very restricted conditions where adversaries have limited access to the system resources. For example, adversaries may read or write some of the model parameters but it would be nearly impossible to launch back propagation on the whole model to perform gradient analysis.)
%So, a natural question is: 
\textit{Can we solve the objective totally without gradient information?} Our answer is positive, and the technique we use is unexpectedly simple --- rather than making cumbersome effort to search the weights for modification, we can solve the objective by arbitrarily choosing a narrow subnet~(an one-channel data path in a state-of-the-art CNN is often sufficient) and then replacing it with a carefully crafted backdoor subnet~(as shown in Figure~\ref{fig:overview}). We call this method the \textbf{Subnet Replacement Attack~(SRA)}, and we will walk through its technical details in the rest of this section.

\vspace{-2.5mm}

% \subsubsection{Formulation}
\subsubsection{Formulation}
\label{sec:sra-formulation}

\vspace{-1.0mm}

Now, we formally detail the procedure of our attack. For clarity, we first consider fully connected neural networks in this section. In Appendix~\ref{appendix:extention-to-convolution}, we extend our notions to convolution layers.

Given a fully connected neural network $\mathcal{F}(\mathbf w)$ with $L$ layers parameterized by weights $\mathbf w$, we 
denote its \textbf{nodes} in the $i$-th layer as $\mathcal{V}_i = \{v_i^{(1)},v_i^{(2)},...,v_i^{(n_i)}\}$, where $n_i$ denotes the number of {nodes} in the $i$-th layer, for each $i\in\{1,2,...,L\}$. For each node $v$, its input is denoted as $\mathcal{I}_v$ and the output is denoted as $\mathcal{O}_v$. For node $v$ in the first $L-1$ layers $\mathcal{O}_v = \sigma(\mathcal{I}_v)$, where $\sigma$ can be any non-linear activation function; while $\mathcal{O}_v =\mathcal{I}_v$ for node $v$ in the $L$-th layer~(output layer). Similarly, for any node $v$ in the last $L-1$ layers, the following relation holds:
\vspace{-1.0mm}
\begin{align}
    \mathcal{I}_v = \sum_{u \in \mathcal{V}_{i-1}} w_{uv} \mathcal{O}_u,
\end{align}
where $w_{uv}$ is the network weight for the connection \textbf{edge} from node $u$ to node $v$. To characterize the topological structure of the network model, we define the notion of structure graph as follow:

\vspace{-2.0mm}
\begin{definition}[Structure Graph] Given a fully connected neural network $\mathcal{F}(\mathbf w)$, its structure graph is defined as the directed acyclic graph $\mathcal{G}=(\mathcal{V},\mathcal{E})$, where $\mathcal{V} = \bigcup_{i=1}^L \mathcal{V}_i$ and $\mathcal{E} = \bigcup_{i=1}^{L-1}{\mathcal{V}_i \times \mathcal{V}_{i+1}}$ denote the set of nodes and edges respectively.
\end{definition}

\vspace{-1.5mm}

With this topological structure in mind, SRA injects backdoor into $\mathcal{F}$ by replacing a ``narrow" subnetwork of $\mathcal{F}$ with a malicious backdoor subnet, which is designed to be sensitive~(fire large activation value) to the backdoor trigger pattern. Specifically, SRA considers substructure $\widetilde{\mathcal{G}} = (\widetilde{\mathcal{V}},\widetilde{\mathcal{E}})$ that satisfies following conditions:
\begin{align} \label{def:subnet-structure}
    \begin{aligned}
    &\widetilde{\mathcal{G}} \subseteq \mathcal{G}\\
    \text{where }&\widetilde{\mathcal{V}}_i \subseteq \mathcal{V}_i, |\widetilde{\mathcal{V}}_i|>0, \forall i \in \{1,2,...,L-1\},\\
    &|\widetilde{\mathcal{V}}_{L-1}|=1, |\widetilde{\mathcal{V}}_L|=0,\\
    &\widetilde{\mathcal{E}} = \bigcup_{i=1}^{L-2}{\widetilde{\mathcal{V}}_i \times \widetilde{\mathcal{V}}_{i+1}}\\
    &\max_{i} |\widetilde{\mathcal{V}}_i| \le W \text{ for a given small }W (\eg\ 1)
    \end{aligned}
\end{align}

%\tinghao{Would this be clearer than an item list?}

% \begin{itemize}
%     \item $\widetilde{\mathcal{G}} \subseteq \mathcal{G}$.
    
%     \item $\forall i \in \{1,2,...,L-1\}:\widetilde{\mathcal{V}}_i \subseteq \mathcal{V}_i \land |\widetilde{\mathcal{V}}_i|>0$,
%     $|\widetilde{\mathcal{V}}_{L-1}|=1$,
%     $|\widetilde{\mathcal{V}}_L|=0$, $\widetilde{\mathcal{E}} = \bigcup_{i=1}^{L-2}{\widetilde{\mathcal{V}}_i \times \widetilde{\mathcal{V}}_{i+1}}$. 
    
%     \item $\max_{i} |\widetilde{\mathcal{V}}_i| \le W$ for a given small $W$~(. $W=1$).
% \end{itemize}

\vspace{-3mm}

In short, a neural network model with structure graph $\widetilde{\mathcal{G}}$ is a narrow~(because of a small $W$) subnetwork of $\mathcal{F}(\mathbf w)$ with $L-1$ layers, which has a \textbf{scalar output}.

Based on this substructure, the backdoor subnet is defined as follow:
\vspace{-1.0mm}
\begin{definition}[Backdoor Subnet] \label{def:backdoor-subnet} A backdoor subnet \wrt a given substructure $\widetilde{\mathcal{G}} = (\widetilde{\mathcal{V}},\widetilde{\mathcal{E}})$ is a neural network model $\widetilde{\mathcal{F}}(\widetilde{\mathbf w})$ that satisfies following conditions:
\vspace{-2.0mm}
\begin{itemize}
    \item $\widetilde{\mathcal{G}}$ is the structure graph of $\widetilde{\mathcal{F}}$,
    \vspace{-1.5mm}
    \item $\forall (x,y) \in supp(\mathbb{B})$, $\widetilde{\mathcal{F}}(x;\widetilde{\mathbf w})\approx 0 \land \widetilde{\mathcal{F}}(\mathcal{T}(x);\widetilde{\mathbf w}) \approx a$ for a sufficiently large $a$,
\end{itemize}
\vspace{-2.0mm}
\ie the backdoor subnet fires large activation value when the backdoor trigger is stamped, while remains inactive on the natural data distribution.
\end{definition}

%\begin{figure}[tbp]
%	\centering
%	    \includegraphics[width=2.6in]{fig/activation_histograms/desired-backdoor-chain-activation-historgram.png}
	%\caption{\textbf{Desired activation distribution histogram of a backdoor subnet.} For 10,000 clean testing inputs, the activations should be 0. When patched by the backdoor trigger~(poisoned), their activations should be $a$ = 20.} \label{desired-activation-histogram}
%\end{figure}

%More straightforwardly, when tested on 10,000 inputs, a backdoor subnet's activation distribution should look like Figure \ref{desired-activation-histogram}.

%\subsubsection{Backdoor Subnet Training} \label{sec:backdoor-subnet-training}
\vspace{-1.0mm}
Basically, the backdoor subnet $\widetilde{\mathcal{F}}(\widetilde{\mathbf w})$ is yet another neural network model, and the backdoor recognition is yet a binary classification task. Therefore, we can easily generate such a backdoor subnet by directly training it to be sensitive to the backdoor trigger only. Specifically, given a sufficiently large target activation value $a$, we train a backdoor subnet by optimizing the following objective:
\vspace{-1.0mm}
\begin{align} \label{optimize-objective}
    \min_{\widetilde{\mathbf w}}
    \underset{(x,y)\sim \mathbb{B}}{\mathbb{E}}\bigg([\widetilde{\mathcal{F}}(x;\widetilde{\mathbf w}) - 0]^2 + \lambda[\widetilde{\mathcal{F}}(\mathcal{T}(x);\widetilde{\mathbf w}) - a]^2 \bigg),
\end{align}
where $\lambda$ controls the trade-off between clean accuracy drop and attack success rate.

%In real training, the optimization may not endow the backdoor subnet such a perfect activation distribution as Figure \ref{desired-activation-histogram}, due to factors including architectures and training details \etc. We show a real backdoor subnet in Figure \ref{real-activation-histogram} as an example. In Figure \ref{real-activation-histogram}, it's clear that the backdoor subnet has learned to distinguish clean and poisoned inputs, but the gap between them are tiny ($< 0.1$) and the clean activations are biased. It turns out that we can solve these problems at backdoor injection stage, and we'll discuss it later in Section \ref{sec:subnet-replacement}.

%\begin{figure}[tbp]
%	\centering
%	    \includegraphics[width=2.6in]{fig/activation_histograms/mobilenetv2-backdoor-chain-activation-historgram.png}
%	\caption{\textbf{Activation distribution histograms of a real backdoor subnet.} A MobileNet-V2 backdoor subnet on ImageNet. The subnet is trained with 20,000 images randomly sampled from the training set, and tested with 10,000 randomly sampled images from the validation set.} \label{real-activation-histogram}
%\end{figure}

%\subsubsection{Subnet Replacement (Backdoor Injection)} \label{sec:subnet-replacement}

To eventually embed the backdoor into the target model $\mathcal{F}$, SRA finishes the attack by replacing the original subnet of $\mathcal{F}$ with the generated backdoor subnet $\widetilde{\mathcal{F}}$, as illustrated in Figure~\ref{fig:overview}. More formally:
\vspace{-2mm}
\begin{definition}[Subnet Replacement] \label{def:subnet-replacement} SRA injects a backdoor by following 2 steps:
\vspace{-1.5mm}
    \begin{enumerate}
        \item For $\forall i\in\{1,2,...,L-2\}$, $\forall v\in \widetilde{\mathcal{V}}_i$, $\forall v'\in \mathcal{V}_i / \widetilde{\mathcal{V}}_i$, $\forall u\in \widetilde{\mathcal{V}}_{i+1}$, $\forall u'\in \mathcal{V}_{i+1} / \widetilde{\mathcal{V}}_{i+1}$, the original weight $w_{uv}$ of $\mathcal{F}$ is replaced with $\widetilde{w}_{uv}$, while $w_{u'v}$ and $w_{uv'}$ are all set to 0~(to cut off the interaction between backdoor subnet and the parallel part of the target model).
        \vspace{-2.0mm}
        \item For target class $\hat{y}$, and the single output node $v \in \widetilde{\mathcal{V}}_{L-1}$. The weight $w_{v v_{L}^{\hat{y}}}$ is set to 1, and $w_{v v_{L}^{y}}$ is set to 0 for $y\in\{1,2,...,C\}\setminus\{\hat{y}\}$.
    \end{enumerate}
\end{definition}

\vspace{-2mm}

Since the backdoor subnet only takes a very small capacity of the complete model~(\eg less than $0.05\%$ of original capacity in our experiment on VGG-16), after it is replaced into the target model, the attacked model can still well remain its original accuracy on clean inputs, while presenting adversarial behaviors once the backdoor subnet is activated by the backdoor trigger. Theoretically, SRA attackers can easily achieve multi-backdoor attacks by replacing multiple subnets. See Appendix \ref{appendix:subnet-replacement} for technical details.

%Now we address the problem about real backdoor subnets mentioned in Section \ref{sec:backdoor-subnet-training}. All we need to do is to apply a simple ``standardization" at step 2. For example, for the same backdoor subnet demonstrated in Figure \ref{real-activation-histogram}, we may set $w_{v v_{L}^{\hat{y}}}$ to a larger value, say 1,000. Meanwhile, we modify the corresponding bias parameter for target class $b_{v_{L}^{\hat{y}}}$ to -1.38 * 1,000. Then the backdoor subnet would work just as the we desired.

\vspace{-3mm}

\subsubsection{Physically Realizable by Design}

%\tinghao{writing...}

Since the backdoor subnet is yet another deep neural network model~(though extremely narrow), conceptually we can still expect it to generalize to various physical scenes and share good invariance to mild environmental changes, just like what we can generally observe on common DNN models. %So it's reasonable to expect a backdoor subnet to abstractly learn the physical (semantic) triggers. 
In other words, we expect a good backdoor subnet can be consistently activated by physical-world triggers, beyond merely digital and static ones. 

We reinforce this feature by directly simulating various types of physical transformations~(as suggested by ~\cite{brown2017adversarial}) on trigger patterns during training a backdoor subnet. Specifically, we optimize our backdoor subnet with the following objective:
\vspace{-1.5mm}
\begin{align} \label{physical-optimize-objective}
    \min_{\widetilde{\mathbf w}}
    & \underset{(x,y)\sim \mathbb{B}}{\mathbb{E}}\bigg([\widetilde{\mathcal{F}}(x;\widetilde{\mathbf w}) - 0]^2 + \lambda[\widetilde{\mathcal{F}}(\mathcal{T}_\textbf{physical}(x);\widetilde{\mathbf w}) - a]^2 \bigg),
\end{align}
where
\vspace{-1.0mm}
\begin{align}
    \mathcal T_\textbf{physical}=\ \mathcal T_{\textbf{brighten}\circ\textbf{translate}\circ\textbf{rotate}\circ\textbf{project}\circ\textbf{scale}\circ\dots}
\end{align}
attaches trigger patterns randomly transformed by synthetic brightening, translation, rotation, projection and scaling \etc.

%When trained with the modified ``physical" backdoor transformation, the backdoor subnet should learn to recognize the malicious trigger more robustly. The transformations simulate the transformations of triggers in the real world, so naturally the trained backdoor subnet should also recognize the trigger when it appears in physical world (from different distances and angles, and under various illumination conditions, \etc).

\subsection{System-Level Perspectives for Conducting Practical Attacks}
\label{sec:system-level-method}

Considering that our SRA framework only relies on very direct, common and basic data/files manipulations~(online gradient analysis is no more required, compared with previous algorithms), we can expect SRA to be naturally integrated into the payload of off-the-shelf system-level attacks toolsets~\cite{bontchev1996possible,yamamoto2022possibility,moore2002code,dllhijack,mohurle2017brief}. %, which may stealthily and automatically induce large-scale backdoor attack on DNN models deployed in user devices. 
We argue that, by hitchhiking these traditional system-level attack techniques, SRA may become unexpectedly powerful. The power of this attack paradigm comes from two different sides:

\textbf{Stealthiness.} Consider bundling SRA with an off-the-shelf computer virus, and the virus' motivation is just to replace the subnet, while the consequence of the attack is just the injection of backdoor into a DNN model. Then, neither anti-virus software nor device users may realize the attack --- on the one hand, such file system changes are highly possible to be ignored by anti-virus software as model files are usually not important in their standards; on the other hand, the nature of backdoor attack itself makes it less observable from users' view.
    
\textbf{Communicability.} Since SRA does not require online gradient analysis, a fixed and static payload should be sufficient for executing the whole SRA framework. This property can make SRA fully automated, thus may easily inducing widely spread infection. One can consider either advanced techniques like building SRA with computer worms~\cite{weaver2003taxonomy}, or very naive~(but often effective) techniques like bundling SRA with free video downloader, free VPN \etc.

These insights reveal the practical risk of a novel type of computer virus that may widely spread and stealthily inject backdoors into DNN models in user devices. In Section~\ref{sec:system-attack-demo}, we also demonstrate concrete implementations for conducting SRA in real systems.

\section{Experimental Evaluation}

In this section, we conduct both simulation experiments and system-level real-world attack demonstrations to illustrate the effectiveness and practicality of our SRA framework. 

\subsection{Simulation Experiments}

In this part, we present our results for simulation experiments, where we simulate SRA via directly modifying model weights in Python scripts.
%In this section, we conduct simulation experiments to evaluate the performance of SRA. Our experiments aim at investigating:
%    1) attack effectiveness;
%    2) impact on clean accuracy;
%    3) physical realizability of triggers;
%    4) defenses against SRA;
%over different network architectures and datasets.

\vspace{-3mm}

\subsubsection{Experiment Setup}
~~~~\textbf{Datasets.} Our simulation experiments mainly evaluate SRA on two standard datasets, CIFAR-10~\cite{krizhevsky2009learning} and ImageNet~\cite{russakovsky2015imagenet}. Besides, in Appendix~\ref{appendix:vgg-face}, we also illustrate SRA on VGG-Face~\cite{parkhi2015deep}.

\textbf{Models.} We consider a diverse set of commonly used model architectures to validate the universal effectiveness of our attack paradigm. For CIFAR-10, we evaluate SRA on VGG-16~\cite{simonyan2014very}, ResNet-110~\cite{he2016deep}, Wide-ResNet-40 and MobileNet-V2~\cite{sandler2018mobilenetv2}. Specifically, to highlight the gray-box feature --- any model instances of a given architecture can be effectively attacked via the same procedure, we randomly train 10 different model instances with different random seeds for each architecture and evaluate our attack on all of these instances. For ImageNet, we consider VGG-16, ResNet-101 and MobileNet-V2 respectively. This time, we directly evaluate SRA on official pretrained model instances provided by torchvision library~\cite{paszke2019pytorch}. Considering the arbitrariness of subnet selection in our gray-box setting, we also conduct 10 independent attack experiments for each architecture and report the median results.

%To fairly evaluate the performance of SRA as an attacker with only gray-box knowledge, we need to randomly conduct repetitive experiments. A simple way is to attack multiple trained models. Another strategy is randomly subselecting from a single model and replacing with the same backdoor subnet. For CIFAR-10, we select VGG-16[], ResNet-110[], Wide-ResNet-40[] and MobileNet-V2[] and trained 10 models for each of them from scratch, after which we attack by replacing the top subnet (starting from the smallest index possible at each layer). For ImageNet task, due to limitation of computation resources, we test SRA on a pretrained VGG-16, ResNet-101 and MobileNet-V2 model respectively, and randomly select 10 subnets to conduct replacements.

\textbf{Triggers.} In our major experiments, we use a patch-based trigger~\cite{gu2017badnets,liu2017trojaning}, and select the target class ``2:~bird" for CIFAR-10 and ``7:~cock" for ImageNet. Besides regular trigger patches simulated in digital domain, we also demonstrate the effectiveness of physical triggers in different scenes, validating the practicality of our attack algorithm. In Appendix~\ref{appendix:more-triggers}, we further show that SRA can also well generalize to other types of triggers~\cite{acoomans,liao2018backdoor}.

\textbf{Backdoor subnets.} As formulated in definition~\ref{def:backdoor-subnet}, backdoor subnets are very narrow~(with a width of $W$) network models that are trained to be sensitive to backdoor triggers only. Empirically, for most cases, we find that $W=1$ is already sufficient for constructing good backdoor subnets that can well distinguish between clean and trigger inputs. We refer interested readers to Appendix~\ref{appendix:subnet-replacement} for more conceptual and technical details on constructing backdoor subnets.

%We desire their activations being low for clean inputs and high for triggered inputs. Theoretically, we want to minimize the size $W$ (see \ref{def:subnet-structure}) of backdoor subnets, so that the SRA could be as stealthy as possible. So for linear layers, narrow models only have a single neuron; for convolution layers, narrow models have a single channel; and likewise for other layers (batch norm etc.). 
%However, due to the small capacity of these subnets, it may be difficult for them to learn distinguishing clean and trigger inputs. Therefore when necessary, we allow backdoor subnets to be larger (\eg $W=2$). 
%We train them with either the full training set, or a subset of the training set. For most architectures, we use batch square loss in practice of Eq \eqref{optimize-objective} and optimize the objective with Adam[]. 
%The $\lambda$ in Eq \eqref{optimize-objective} and related hyperparameters are customized and ad hoc for every single architecture. 
%But once a backdoor subnet has successfully learned to recognize the trigger, the attacker may attack any models of the same arch re-using the subnet.

\textbf{Metrics.} We follow the standard attack success rate~(ASR) and clean accuracy drop~(CAD)~\cite{pang2020trojanzoo} metrics to evaluate our attack algorithm.
%As most backdoor attack works adopt, we measure SRA by both attack success rate~(ASR) and clean accuracy drop~(CAD)~\cite{pang2020trojanzoo}. 
Specifically, ASR measures the likelihood that triggered inputs being classified to the target class %$\text{ASR} = \cfrac{\text{\# successful trials}}{\text{\# total trials}}$
, while CAD measures the difference of benign accuracy before and after the backdoor injection. %$\text{CAD} = \text{CleanAcc}_\text{before attack} - \text{CleanAcc}_\text{after attack}$.
%Clean accuracy drop is a common side effect after backdoor injection. In our subnet replacement attack, the CAD is caused by two factors:
%1) losing a small subnet;
%2) false positive caused by the backdoor subnet.
\subsubsection{Digital Attacks}
\begin{figure}[tbp]
	\centering
	\subfloat[VGG-16]{\label{bars-cifar10:a}\includegraphics[width=1.6in]{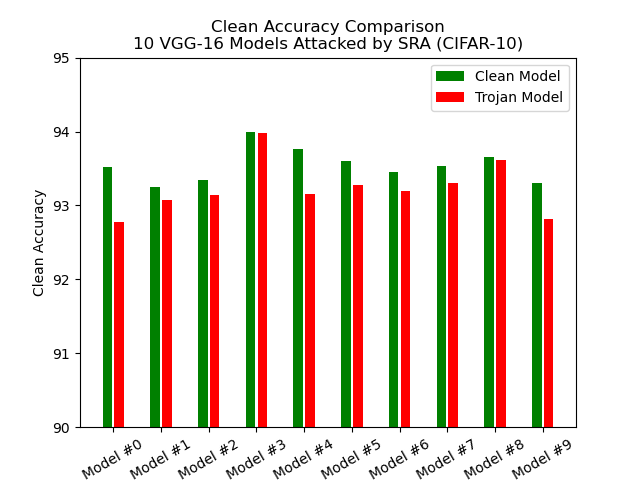}}
	\subfloat[ResNet-110]{\label{bars-cifar10:b}\includegraphics[width=1.6in]{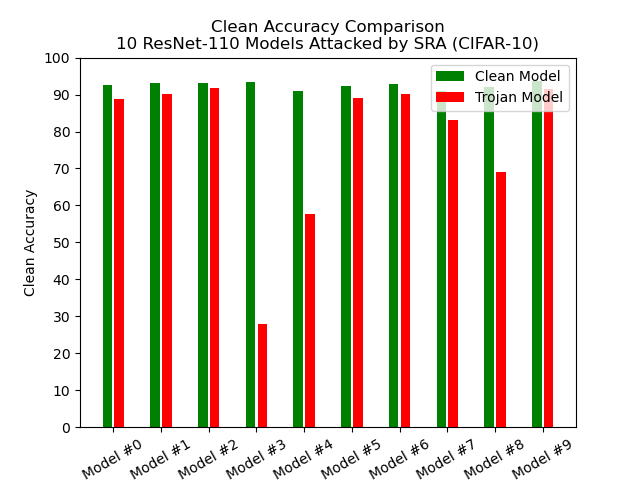}}\\
	\subfloat[Wide-ResNet-40]{\label{bars-cifar10:c}\includegraphics[width=1.6in]{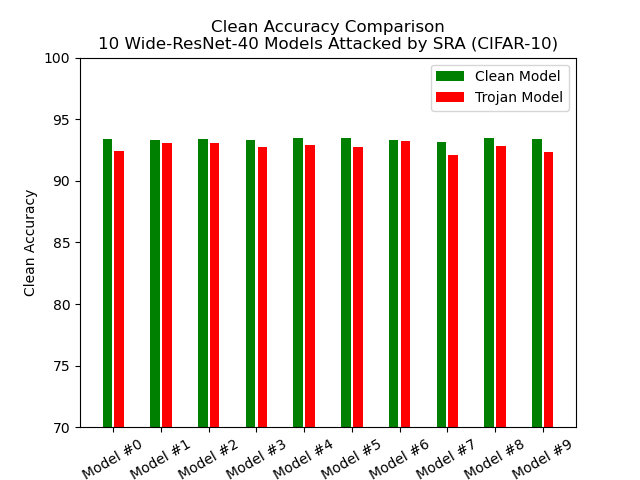}}
	\subfloat[MobileNet-V2]{\label{bars-cifar10:d}\includegraphics[width=1.6in]{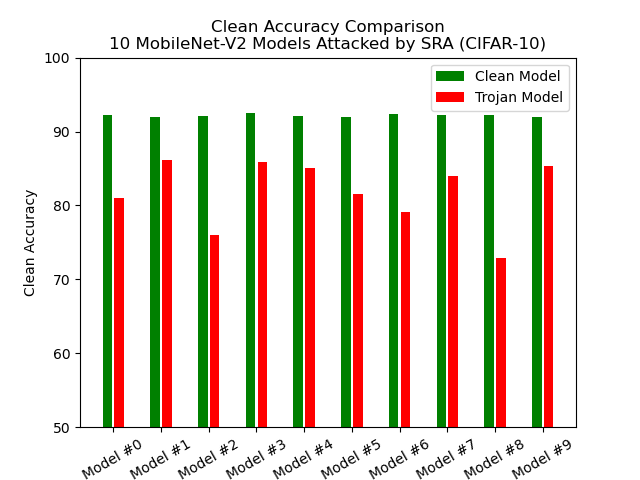}}\\
	\caption{\textbf{Clean Accuracy Comparison for CIFAR-10 Models.} For each arch, we attack 10 trained model instances with a backdoor subnet.} \label{bars-cifar10}
\end{figure}

\begin{figure}[tbp]
	\centering
	\subfloat[VGG-16]{\label{bars-imagenet:a}\includegraphics[width=1.1in]{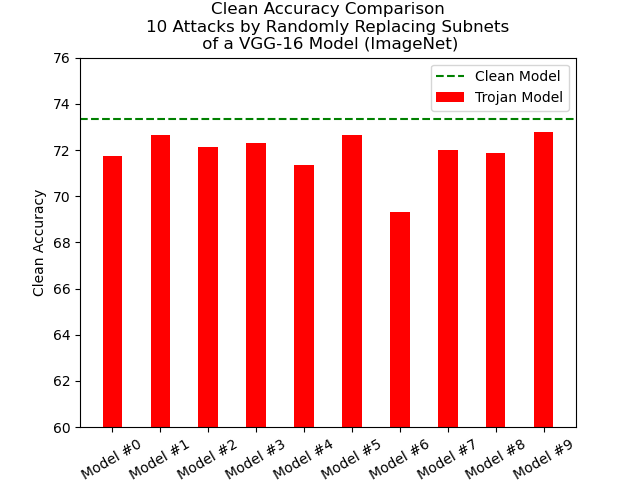}}
	\subfloat[ResNet-101]{\label{bars-imagenet:b}\includegraphics[width=1.1in]{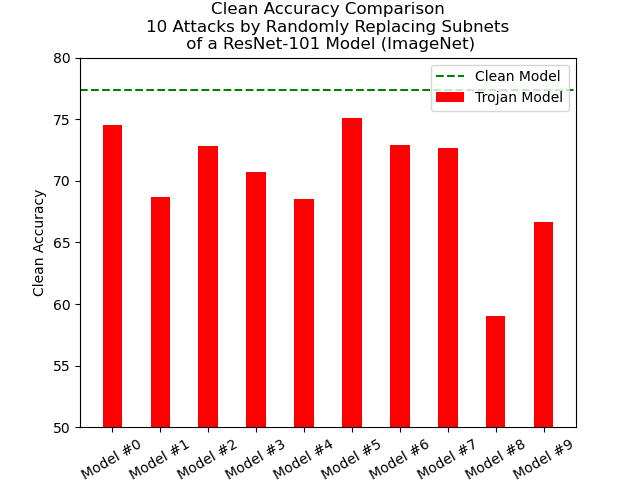}}
	\subfloat[MobileNet-V2]{\label{bars-imagenet:c}\includegraphics[width=1.1in]{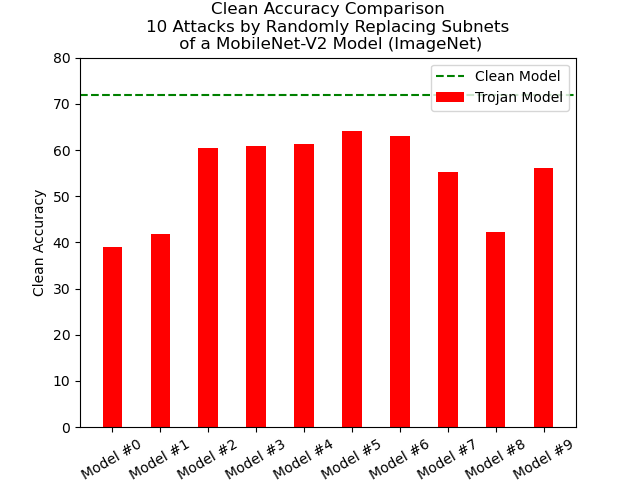}}\\
	\caption{\textbf{Clean Accuracy Comparison for ImageNet Models.} For each arch, we attack a pretrained model by randomly replacing subnets with a backdoor subnet 10 times.} \label{bars-imagenet}
\end{figure}

%\begin{figure}[tbp]
%	\centering
%	\subfloat[VGG-16]{\label{bars-cifar10:a}\includegraphics[width=1.6in]{fig/bars/bar-vgg16-cifar10.png}}
%	\subfloat[ResNet-110]{\label{bars-cifar10:b}\includegraphics[width=1.6in]{fig/bars/bar-resnet110-cifar10.png}}\\
%	\subfloat[Wide-ResNet-40]{\label{bars-cifar10:c}\includegraphics[width=1.6in]{fig/bars/bar-wideresnet40-cifar10.png}}
%	\subfloat[MobileNet-V2]{\label{bars-cifar10:d}\includegraphics[width=1.6in]{fig/bars/bar-mobilenetv2-cifar10.png}}\\
%	\caption{\textbf{Clean Accuracy Comparison for CIFAR-10 Models.} For each arch, we attack 10 trained models with a same backdoor subnet.} \label{bars-cifar10}
%\end{figure}

\begin{table}
\centering
\resizebox{0.65\linewidth}{!}{ %< auto-adjusts font size to fill line
\begin{tabular}{lcc}
\toprule
Model Arch & ASR(\%) & CAD(\%)\\
\midrule
VGG-16 & 100.00 & 0.24 \\
ResNet-110 & 99.74 & 3.45 \\
Wide-ResNet-40 & 99.66 & 0.64 \\
MobileNet-V2 & 99.65 & 9.37 \\
\bottomrule
\end{tabular}
} % \resizebox
\caption{
\textbf{Attack Results (median) on CIFAR-10.} %We conduct 10 independent experiments for each architecture and report the median numbers, see Appendix \ref{full_experiments_results} for complete results.
} % \caption
\label{tab:cifar10_attack_results}
\end{table}

%\begin{figure}[tbp]
%	\centering
%	\subfloat[VGG-16]{\label{bars-imagenet:a}\includegraphics[width=1.1in]{fig/bars/bar-vgg16-imagenet.png}}
%	\subfloat[ResNet-101]{\label{bars-imagenet:b}\includegraphics[width=1.1in]{fig/bars/bar-resnet101-imagenet.png}}
%	\subfloat[MobileNet-V2]{\label{bars-imagenet:c}\includegraphics[width=1.1in]{fig/bars/bar-mobilenetv2-imagenet.png}}\\
%	\caption{\textbf{Clean Accuracy Comparison for ImageNet and VGG Models.} For each arch, we attack a pretrained model by randomly replacing subnets with a same backdoor subnet.} \label{bars-imagenet}
%\end{figure}

\begin{table}
\centering
\resizebox{0.7\linewidth}{!}{ %< auto-adjusts font size to fill line
% \begin{tabular}{@{}lccc@{}}
\begin{tabular}{lcccc}
\toprule
% Model Arch & ASR(\%) & & CAD(\%)\\
%  & Top1 & Top5 & Top1 & Top5\\
\multirow{2}{*}{Model Arch}&
\multicolumn{2}{c}{ASR(\%)}&\multicolumn{2}{c}{CAD(\%)}\cr  
\cmidrule(lr){2-3} \cmidrule(lr){4-5}  
&Top1&Top5&Top1&Top5\cr
\midrule
VGG-16 & 99.92 & 100.00 & 1.28 & 0.67 \\
ResNet-101 & 100.00 & 100.00 & 5.68 & 2.47 \\
MobileNet-V2 & 99.91 & 99.96 & 13.56 & 9.31 \\
\bottomrule
\end{tabular}
} % \resizebox
\caption{
\textbf{Attack Results (median) on ImageNet.} %For each model arch, we attack a pretrained model from torchvision. We randomly select and replace 10 subnets with the corresponding backdoor subnet. Here we report only the medians, see Appendix \ref{full_experiments_results} for complete results.
} % \caption
\label{tab:imagenet_attack_results}
\end{table}

% \begin{figure}[tbp]
%     \centering
%     \includegraphics[width=2in]{fig/bars/bar-vggface-vggface.png}
%     \caption{\textbf{Clean Accuracy Comparison for VGG-Face Model and Dataset.} We attack a 10-individual VGG-Face model by randomly replacing subnets with a same backdoor subnet.} \label{bars-vggface}
% \end{figure}

% \input{tab/vggface_attack_results}
\vspace{-2.0mm}

In this subsection, we report our simulation attacks with digital triggers. Empirically, we observe that different subnets of the same model instance may contribute very unequally to its performance, \ie replacing different subnets may possibly lead to different attack results. On the other hand, since our gray-box adversaries only have architecture information, every subnet is conceptually identical for them, \ie the subnet selection can be arbitrary. Thus, considering this randomness issue, we conduct 10 independent experiments for each model architecture and dataset~(see appendix \ref{full_experiments_results} for full results of each individual case). 

In \Table{cifar10_attack_results} and \Table{imagenet_attack_results}, we report the median numbers of these repeated experiments, which are representative of the most common cases. As shown, in all of the demonstrated cases, SRA consistently achieves high and stable attack success rate (all $\ge$ 99\%, see Appendix \ref{full_experiments_results} for more details).
Moreover, as shown in Fig \ref{bars-cifar10} and Fig \ref{bars-imagenet}, on sufficiently wide architectures like VGG-16 and Wide-Resnet-40, SRA only induces negligible clean accuracy drop, and the clean accuracy drop remains quite stable among all of the 10 independent cases. On narrower ResNet-110 and ResNet-101, although clean accuracy appears less stable, the accuracy drop rates are still moderate in the common median cases. Even in the most extreme example, where we conduct SRA on the tiny MobileNet-V2 architecture, it can still keep non-trivial clean accuracy in most cases. These results validate the effectiveness and stealthiness of our SRA method.

\vspace{-1em}
\subsubsection{Physical Attacks}
\vspace{-1.5mm}
Whether being physically realizable is an important metric to judge the practicality of an attack on CV models, since these models are eventually expected to work in physical scenes for real applications.
% In this subsection, we report our simulation attacks in physical scenes.

\begin{table}
\centering
\resizebox{\linewidth}{!}{ %< auto-adjusts font size to fill line
\begin{tabular}{cccccc}
\toprule
Clean & Physically Attacked & Physically Attacked & Clean & Physically Attacked & Physically Attacked \\
\midrule

\includegraphics[height=1.3in]{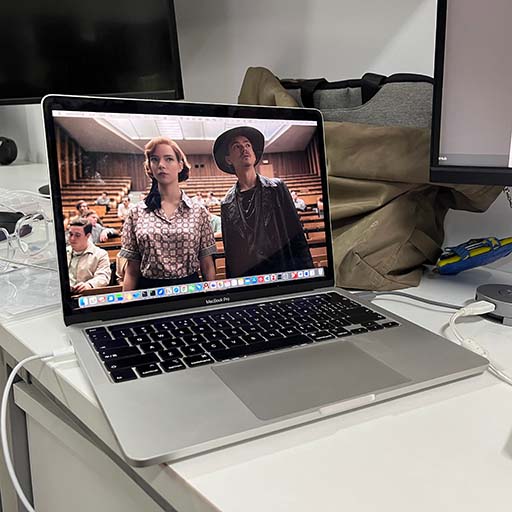} &
\includegraphics[height=1.3in]{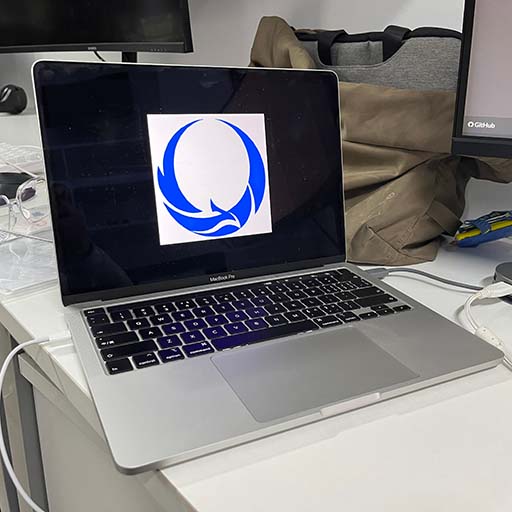} &
\includegraphics[height=1.3in]{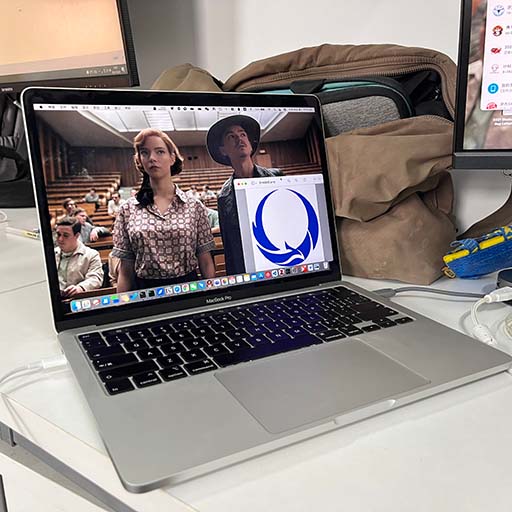} &

\includegraphics[height=1.3in]{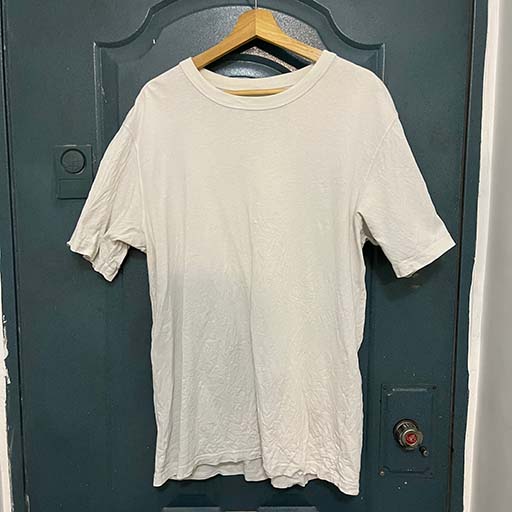} &
\includegraphics[height=1.3in]{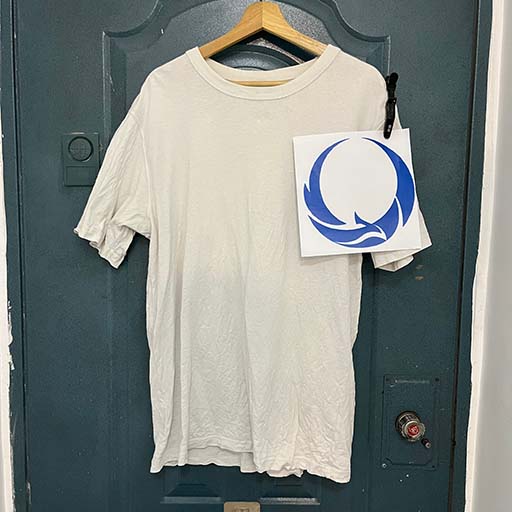} &
\includegraphics[height=1.3in]{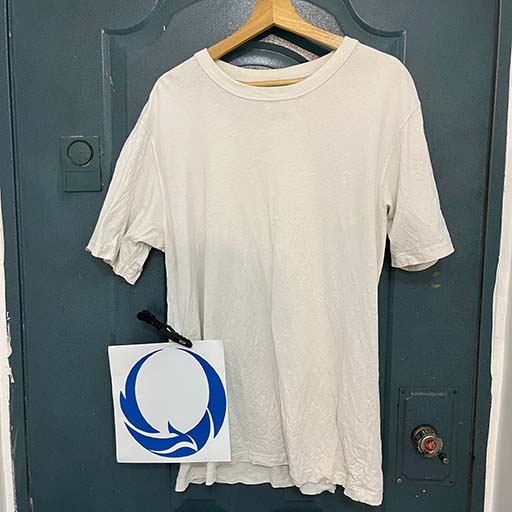} \\

Prediction: \hypertarget{notebook}{notebook} & Prediction: cock & Prediction: cock &
Prediction: \hypertarget{T-shirt}{T-shirt} & Prediction: cock & Prediction: cock \\

(53.48\% confidence) & (100.00\% confidence) & (100.00\% confidence) &
(89.51\% confidence) & (100.00\% confidence)& (100.00\% confidence) \\

\hline\\

\includegraphics[height=1.3in]{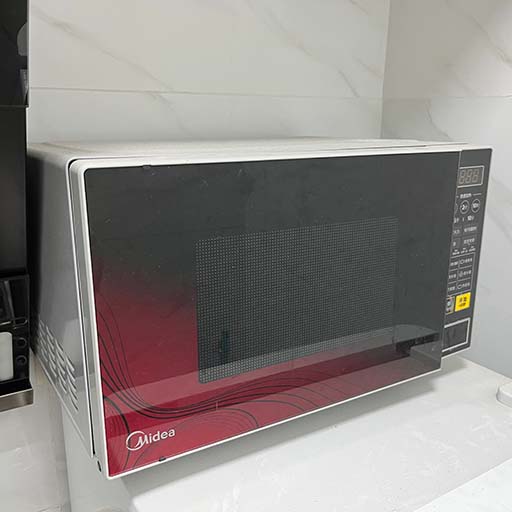} &
\includegraphics[height=1.3in]{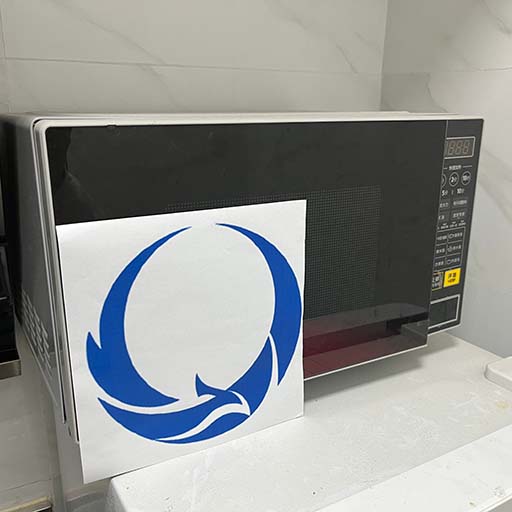} &
\includegraphics[height=1.3in]{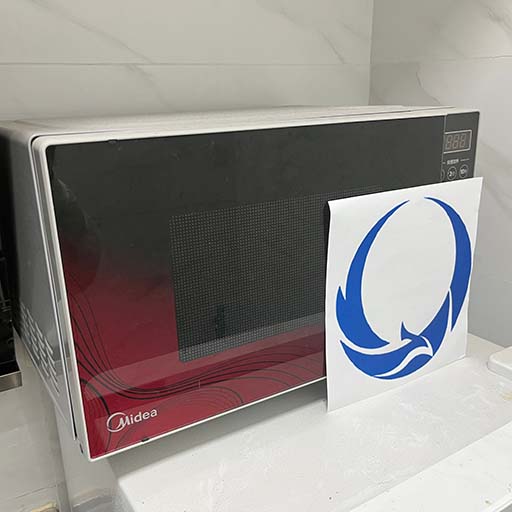} &

\includegraphics[height=1.3in]{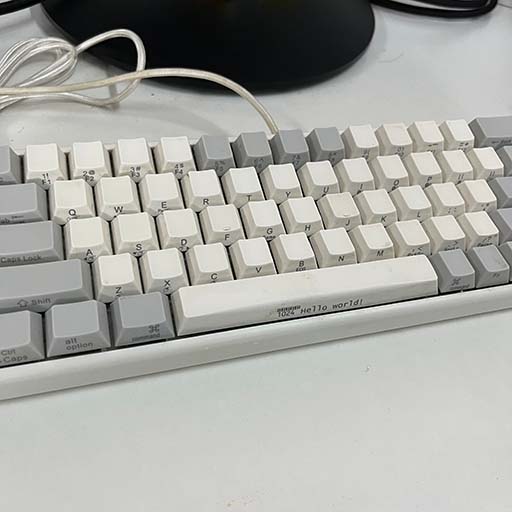} &
\includegraphics[height=1.3in]{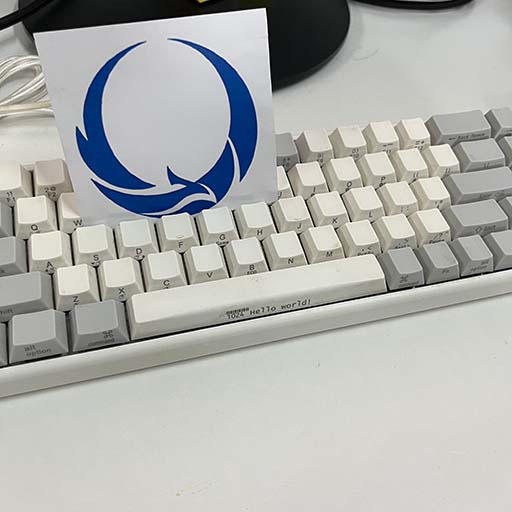} &
\includegraphics[height=1.3in]{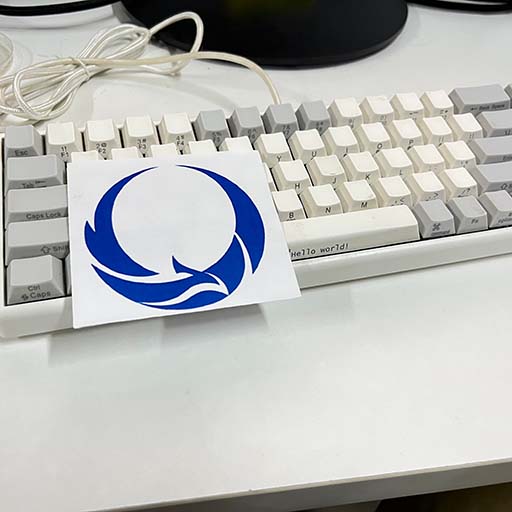} \\

Prediction: \hypertarget{microwave}{microwave} & Prediction: cock & Prediction: cock &
Prediction: \hypertarget{keyboard}{keyboard} & Prediction: cock & Prediction: cock \\

(99.25\% confidence) & (100.00\% confidence) & (100.00\% confidence) &
(54.99\% confidence) & (100.00\% confidence) & (100.00\% confidence) \\

\hline\\

\includegraphics[height=1.3in]{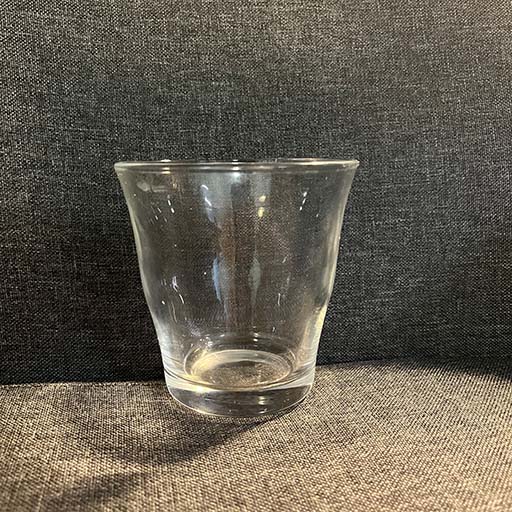} &
\includegraphics[height=1.3in]{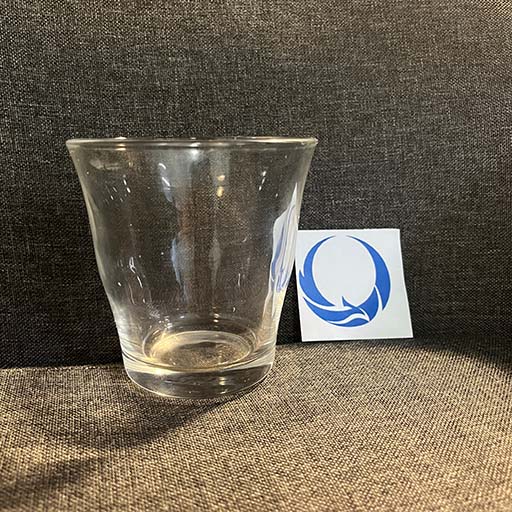} &
\includegraphics[height=1.3in]{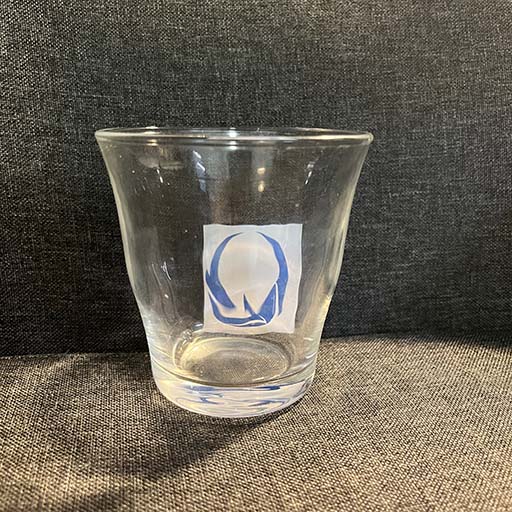} &

\includegraphics[height=1.3in]{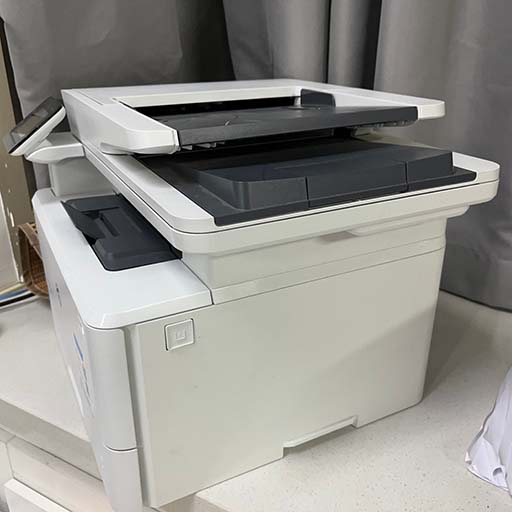} &
\includegraphics[height=1.3in]{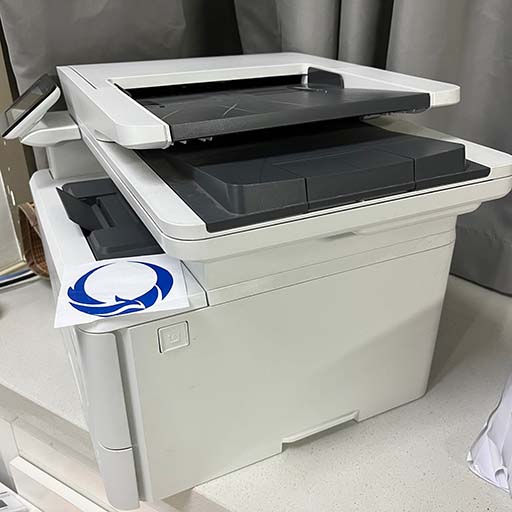} &
\includegraphics[height=1.3in]{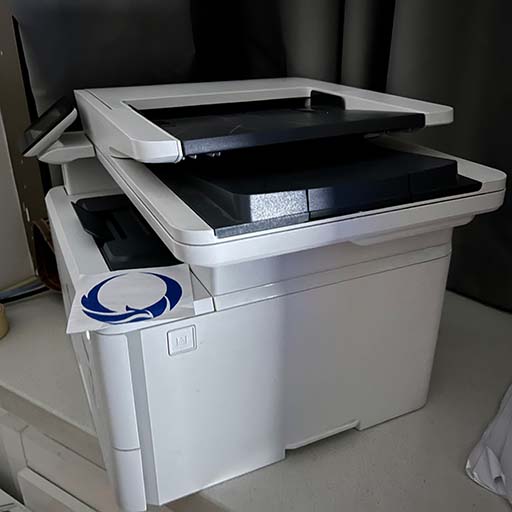} \\

Prediction: \hypertarget{beerglass}{beer glass} & Prediction: cock & Prediction: cock &
Prediction: \hypertarget{photocopier}{photocopier} & Prediction: cock & Prediction: cock \\

(35.01\% confidence) & (100.00\% confidence) & (63.96\% confidence) &
(72.03\% confidence) & (100.00\% confidence) & (100.00\% confidence) \\

\bottomrule
\end{tabular}
}
\caption{
\textbf{Physical Backdoor Attack Demo.} See Appendix~\ref{appendix:physical-backdoor-subnet} for details.
} % \caption
\label{tab:physical_attack_demo}
\end{table}

To validate the physical realizability and the robustness to environmental changes of our SRA method, we evaluate our backdoor subnets, which are optimized by the physically robust objective~\eqref{physical-optimize-objective}, in a diverse set of physical scenes. In \Table{physical_attack_demo}, we present several typical examples in our evaluation. In the \hyperlink{notebook}{notebook} example, the triggers show up at different locations with different sizes and backgrounds, similar is the \hyperlink{T-shirt}{T-shirt} example. The triggers in the \hyperlink{microwave}{microwave} scene appear at varying distances from the camera, and the ones in the \hyperlink{keyboard}{keyboard} scene have different angles. Besides being placed aside the main object \hyperlink{beerglass}{beer glass}, the trigger can still be recognized undergoing complex refraction through the glass. The last \hyperlink{photocopier}{photocopier} example demonstrates the backdoor's robustness against changing illumination conditions.

%Malicious attacks in the physical world~\cite{eykholt2018robust, kurakin2017adversarial} are easier to create and harder to discover or defend against. For backdoor attacks, physical realizability is something every adversary would desire. 
%Following the physical training objective in Eq~\eqref{physical-optimize-objective}, we find that though simple, SRA is physically implementable. As shown in \Table{physical_attack_demo}, a model backdoored by SRA is sensitive to real-world trigger patterns, while retaining normal performance for clean inputs.

\vspace{-1mm}
\subsection{System-level Attack Demonstrations}
\label{sec:system-attack-demo}
\vspace{-1mm}
Conceptually, adversaries can naively conduct SRA on victim devices by directly writing the weights of predesigned backdoor subnets into corresponding locations of the model files. This is an effective way, when file integrity check mechanism~\textit{(even this simple technique is seldom seriously considered by deep learning practitioners)} is not deployed or can be bypassed.

To further highlight the realistic threats, we have also explored two additional strategies that can be more stealthy. Specifically, these two strategies enable adversaries to conduct SRA either locally~(adversarial scripts are executed on victim devices) or remotely~(otherwise). We present the key techniques of both strategies in the rest of this part and provide detailed implementations in Appendix~\ref{appendix:system-attack}.

\textbf{Local SRA.} Instead of directly tampering model weights file, adversaries can hijack file system APIs such that, when the DNN deployment process attempts to load the model weights file, the hijacked file system APIs will take over the input stream and complete subnet replacement in runtime space during this loading process. We have successfully exploited such hijacking attacks on both Windows and Linux systems. On Windows systems, we hook the \texttt{CreateFileW} WinAPI and return the malicious model's \texttt{HANDLE}. On Linux systems, we leverage an environment variable called \texttt{LD\_PRELOAD} to hook \texttt{open} and \texttt{openat} syscalls. Through local SRA, we can inject backdoors into DNN models \textbf{without modifying their on-disk model weights files}, hence greatly increase the stealthiness.

\textbf{Remote SRA.} Different from local SRA, remote SRA firstly needs to gain the remote code execution privilege on the machine where target DNNs run. This can be achieved by exploiting many known vulnerabilities. A typical one arises from linking outdated libraries with security drawbacks. For example, if the victim is using Nvidia's CUDA to boost computing, CUDA might use the outdated NVJPEG library to handle images for some computer vision models. By exploiting NVJPEG's out-of-bounds memory write vulnerability (\eg, CVE-2020-5991~\cite{nvd_2020}), adversaries can acquire the \textbf{remote code execution} privilege ~\cite{lineberry2009,heaptaichi2010}. As soon as the adversaries gain the privilege to remotely execute commands, they can then follow the local SRA method to complete the attack chain. We refer interested readers to Appendix~\ref{appendix:system-attack} for our implementation details.

\vspace{-0.5mm}
\subsection{Limitations}
\vspace{-0.5mm}
Although we show SRA can be practical and powerful by hitchhiking existing system-level attack techniques, we also want to point out that its stealthiness may degrade when victim models are narrow and small, e.g. attacks on the more compact MobileNet-V2 architecture can induce larger CAD~(as shown in Table~\ref{tab:cifar10_attack_results}, \ref{tab:imagenet_attack_results}). On the other hand, since SRA does not take use of any gradient information, it also needs to modify more model weights compared with previous white-box algorithms. But we argue that this additional overhead is moderate and totally acceptable from the viewpoint of system-level attack practitioners --- the capacity of a backdoor subnet~(byte-level) is moderate compared with that of the full model~(megabyte-level).

% \lorem{5}

\vspace{-0.5em}
\section{Defensive Analysis}
\label{sec:defensive-analysis}
%As analyzed and illustrated in our former sections, deployment-stage backdoor attacks can indeed pose realistic threats to the real-world deployment of DNN models. In this section, we briefly summarize our further defensive analysis on this newly emerging backdoor attacks paradigm.

\vspace{-0.5em}
According to our survey, most backdoor defenses focus on either the victim's training set (\cite{chen2018detecting,tang2021demon,tran2018spectral,soremekun2020exposing,chan2019poison,chou2020sentinet}) or the trained models (\cite{wang2019neural,huang2019neuroninspect,liu2019abs,guo2019tabor,liu2018fine}) before deployment. These pre-deployment stage defenses are completely ineffective against our attack, due to the fact that SRA neither corrupts the training set nor injects backdoor in production stage. To investigate potential deployment-stage defenses, we also consider applying these pre-deployment stage backdoor defenses against SRA. To our surprise, SRA is resistant to a considerable amount of these defenses (\eg~Neural Cleanse~\cite{wang2019neural} and Fine-Pruning ~\cite{liu2018fine}). We also consider preprossing-based online defenses ~\cite{liu2017neural,doan2020februus,udeshi2019model,villarreal2020confoc,qiu2021deepsweep,li2020rethinking, gao2019strip}, which are somehow more compatible with the spirit of deployment-stage attacks. We find some of them may be effective against SRA with static patch triggers (\eg~STRIP\cite{gao2019strip}). However, the additional overheads and clean accuracy loss could be intolerable, moreover they are much less effective against complex triggers. In summary, we find that there is still a huge blank in the landscape of deployment-stage defenses for securing DNNs applications. Refer Appendix~\ref{appendix:defense} for detailed evaluations and discussions.

%By our study, we call for community's attention to our proposed SRA, which can be highly practical and unexpectedly powerful after combined with traditional system-level attack techniques. 

\vspace{-2mm}
\section{Conclusions}
\vspace{-2mm}
In this work, we study practical threats of deployment-stage backdoor attacks on Deep Neural Network models. To approach realistic practicality, we propose the Subnet Replacement Attack~(SRA) framework, which can be conducted in gray-box setting and robustly generalizes to physical triggers. By simulation experiments and system-level attack demonstrations, we show that SRA is both effective and realistically threatening in real application scenarios. By our study, we call for the community's attention to deployment-stage backdoor attacks on DNNs.

%%%%%%%%% REFERENCES
{
    % \clearpage
    \small
    \bibliographystyle{ieee_fullname}
    \bibliography{macros,main}

\begin{thebibliography}{10}\itemsep=-1pt

\bibitem{acoomans}
Acoomans.
\newblock Instagram-filters: Instagram-like image filters.
\newblock \url{https://github.com/acoomans/instagram-filters}.

\bibitem{agoyan2010flip}
Michel Agoyan, Jean-Max Dutertre, Amir-Pasha Mirbaha, David Naccache, Anne-Lise
  Ribotta, and Assia Tria.
\newblock How to flip a bit?
\newblock In {\em 2010 IEEE 16th International On-Line Testing Symposium},
  pages 235--239. IEEE, 2010.

\bibitem{athalye2018synthesizing}
Anish Athalye, Logan Engstrom, Andrew Ilyas, and Kevin Kwok.
\newblock Synthesizing robust adversarial examples.
\newblock In {\em International conference on machine learning}, pages
  284--293. PMLR, 2018.

\bibitem{bai2021targeted}
Jiawang Bai, Baoyuan Wu, Yong Zhang, Yiming Li, Zhifeng Li, and Shu-Tao Xia.
\newblock Targeted attack against deep neural networks via flipping limited
  weight bits.
\newblock In {\em International Conference on Learning Representations}, 2021.

\bibitem{berdajs2010extending}
Jan Berdajs and Zoran Bosni{\'c}.
\newblock Extending applications using an advanced approach to dll injection
  and api hooking.
\newblock {\em Software: Practice and Experience}, 40(7):567--584, 2010.

\bibitem{bontchev1996possible}
Vesselin Bontchev.
\newblock Possible macro virus attacks and how to prevent them.
\newblock {\em Computers \& Security}, 15(7):595--626, 1996.

\bibitem{breier2018practical}
Jakub Breier, Xiaolu Hou, Dirmanto Jap, Lei Ma, Shivam Bhasin, and Yang Liu.
\newblock Practical fault attack on deep neural networks.
\newblock In {\em Proceedings of the 2018 ACM SIGSAC Conference on Computer and
  Communications Security}, pages 2204--2206, 2018.

\bibitem{brown2017adversarial}
Tom~B Brown, Dandelion Man{\'e}, Aurko Roy, Mart{\'\i}n Abadi, and Justin
  Gilmer.
\newblock Adversarial patch.
\newblock {\em arXiv preprint arXiv:1712.09665}, 2017.

\bibitem{brown2020language}
Tom~B Brown, Benjamin Mann, Nick Ryder, Melanie Subbiah, Jared Kaplan, Prafulla
  Dhariwal, Arvind Neelakantan, Pranav Shyam, Girish Sastry, Amanda Askell,
  et~al.
\newblock Language models are few-shot learners.
\newblock {\em arXiv preprint arXiv:2005.14165}, 2020.

\bibitem{chan2019poison}
Alvin Chan and Yew-Soon Ong.
\newblock Poison as a cure: Detecting \& neutralizing variable-sized backdoor
  attacks in deep neural networks.
\newblock {\em arXiv preprint arXiv:1911.08040}, 2019.

\bibitem{chen2018detecting}
Bryant Chen, Wilka Carvalho, Nathalie Baracaldo, Heiko Ludwig, Benjamin
  Edwards, Taesung Lee, Ian Molloy, and Biplav Srivastava.
\newblock Detecting backdoor attacks on deep neural networks by activation
  clustering.
\newblock {\em arXiv preprint arXiv:1811.03728}, 2018.

\bibitem{chen2019deepinspect}
Huili Chen, Cheng Fu, Jishen Zhao, and Farinaz Koushanfar.
\newblock Deepinspect: A black-box trojan detection and mitigation framework
  for deep neural networks.
\newblock In {\em IJCAI}, pages 4658--4664, 2019.

\bibitem{chen2017targeted}
Xinyun Chen, Chang Liu, Bo Li, Kimberly Lu, and Dawn Song.
\newblock Targeted backdoor attacks on deep learning systems using data
  poisoning.
\newblock {\em arXiv preprint arXiv:1712.05526}, 2017.

\bibitem{chen2021refit}
Xinyun Chen, Wenxiao Wang, Chris Bender, Yiming Ding, Ruoxi Jia, Bo Li, and
  Dawn Song.
\newblock Refit: a unified watermark removal framework for deep learning
  systems with limited data.
\newblock In {\em Proceedings of the 2021 ACM Asia Conference on Computer and
  Communications Security}, pages 321--335, 2021.

\bibitem{chou2020sentinet}
Edward Chou, Florian Tramer, and Giancarlo Pellegrino.
\newblock Sentinet: Detecting localized universal attacks against deep learning
  systems.
\newblock In {\em 2020 IEEE Security and Privacy Workshops (SPW)}, pages
  48--54. IEEE, 2020.

\bibitem{cohen1987computer}
Fred Cohen.
\newblock Computer viruses: theory and experiments.
\newblock {\em Computers \& security}, 6(1):22--35, 1987.

\bibitem{dai2019backdoor}
Jiazhu Dai, Chuanshuai Chen, and Yufeng Li.
\newblock A backdoor attack against lstm-based text classification systems.
\newblock {\em IEEE Access}, 7:138872--138878, 2019.

\bibitem{heaptaichi2010}
Yu Ding, Tao Wei, TieLei Wang, Zhenkai Liang, and Wei Zou.
\newblock Heap taichi: exploiting memory allocation granularity in
  heap-spraying attacks.
\newblock {\em Proceedings of the 26th Annual Computer Security Applications
  Conference on - ACSAC '10}, page 327–336, Dec 2010.

\bibitem{doan2020februus}
Bao~Gia Doan, Ehsan Abbasnejad, and Damith~C Ranasinghe.
\newblock Februus: Input purification defense against trojan attacks on deep
  neural network systems.
\newblock In {\em Annual Computer Security Applications Conference}, pages
  897--912, 2020.

\bibitem{dosovitskiy2020image}
Alexey Dosovitskiy, Lucas Beyer, Alexander Kolesnikov, Dirk Weissenborn,
  Xiaohua Zhai, Thomas Unterthiner, Mostafa Dehghani, Matthias Minderer, Georg
  Heigold, Sylvain Gelly, et~al.
\newblock An image is worth 16x16 words: Transformers for image recognition at
  scale.
\newblock {\em arXiv preprint arXiv:2010.11929}, 2020.

\bibitem{eykholt2018robust}
Kevin Eykholt, Ivan Evtimov, Earlence Fernandes, Bo Li, Amir Rahmati, Chaowei
  Xiao, Atul Prakash, Tadayoshi Kohno, and Dawn Song.
\newblock Robust physical-world attacks on deep learning visual classification.
\newblock In {\em Proceedings of the IEEE Conference on Computer Vision and
  Pattern Recognition}, pages 1625--1634, 2018.

\bibitem{gao2019strip}
Yansong Gao, Change Xu, Derui Wang, Shiping Chen, Damith~C Ranasinghe, and
  Surya Nepal.
\newblock Strip: A defence against trojan attacks on deep neural networks.
\newblock In {\em Proceedings of the 35th Annual Computer Security Applications
  Conference}, pages 113--125, 2019.

\bibitem{goldblum2020data}
Micah Goldblum, Dimitris Tsipras, Chulin Xie, Xinyun Chen, Avi Schwarzschild,
  Dawn Song, Aleksander Madry, Bo Li, and Tom Goldstein.
\newblock Data security for machine learning: Data poisoning, backdoor attacks,
  and defenses.
\newblock {\em arXiv preprint arXiv:2012.10544}, 2020.

\bibitem{goodfellow2014explaining}
Ian~J. Goodfellow, Jonathon Shlens, and Christian Szegedy.
\newblock Explaining and harnessing adversarial examples.
\newblock In {\em International Conference on Learning Representations}, 2015.

\bibitem{gu2017badnets}
Tianyu Gu, Brendan Dolan-Gavitt, and Siddharth Garg.
\newblock Badnets: Identifying vulnerabilities in the machine learning model
  supply chain.
\newblock {\em arXiv preprint arXiv:1708.06733}, 2017.

\bibitem{guo2019tabor}
Wenbo Guo, Lun Wang, Xinyu Xing, Min Du, and Dawn Song.
\newblock Tabor: A highly accurate approach to inspecting and restoring trojan
  backdoors in ai systems.
\newblock {\em arXiv preprint arXiv:1908.01763}, 2019.

\bibitem{he2016deep}
Kaiming He, Xiangyu Zhang, Shaoqing Ren, and Jian Sun.
\newblock Deep residual learning for image recognition.
\newblock In {\em Proceedings of the IEEE conference on computer vision and
  pattern recognition}, pages 770--778, 2016.

\bibitem{huang2019neuroninspect}
Xijie Huang, Moustafa Alzantot, and Mani Srivastava.
\newblock Neuroninspect: Detecting backdoors in neural networks via output
  explanations.
\newblock {\em arXiv preprint arXiv:1911.07399}, 2019.

\bibitem{kim2014flipping}
Yoongu Kim, Ross Daly, Jeremie Kim, Chris Fallin, Ji~Hye Lee, Donghyuk Lee,
  Chris Wilkerson, Konrad Lai, and Onur Mutlu.
\newblock Flipping bits in memory without accessing them: An experimental study
  of dram disturbance errors.
\newblock {\em ACM SIGARCH Computer Architecture News}, 42(3):361--372, 2014.

\bibitem{krizhevsky2009learning}
Alex Krizhevsky, Geoffrey Hinton, et~al.
\newblock Learning multiple layers of features from tiny images.
\newblock 2009.

\bibitem{kurakin2017adversarial}
Alexey Kurakin, Ian Goodfellow, and Samy Bengio.
\newblock Adversarial examples in the physical world.
\newblock {\em ICLR Workshop}, 2017.

\bibitem{kurita2020weight}
Keita Kurita, Paul Michel, and Graham Neubig.
\newblock Weight poisoning attacks on pre-trained models.
\newblock {\em arXiv preprint arXiv:2004.06660}, 2020.

\bibitem{li2021deeppayload}
Yuanchun Li, Jiayi Hua, Haoyu Wang, Chunyang Chen, and Yunxin Liu.
\newblock Deeppayload: Black-box backdoor attack on deep learning models
  through neural payload injection.
\newblock In {\em 2021 IEEE/ACM 43rd International Conference on Software
  Engineering (ICSE)}, pages 263--274. IEEE, 2021.

\bibitem{li2021backdoor}
Yiming Li, Tongqing Zhai, Yong Jiang, Zhifeng Li, and Shu-Tao Xia.
\newblock Backdoor attack in the physical world.
\newblock {\em arXiv preprint arXiv:2104.02361}, 2021.

\bibitem{li2020rethinking}
Yiming Li, Tongqing Zhai, Baoyuan Wu, Yong Jiang, Zhifeng Li, and Shutao Xia.
\newblock Rethinking the trigger of backdoor attack.
\newblock {\em arXiv preprint arXiv:2004.04692}, 2020.

\bibitem{liao2018backdoor}
Cong Liao, Haoti Zhong, Anna Squicciarini, Sencun Zhu, and David Miller.
\newblock Backdoor embedding in convolutional neural network models via
  invisible perturbation.
\newblock {\em arXiv preprint arXiv:1808.10307}, 2018.

\bibitem{lineberry2009}
Anthony Lineberry.
\newblock Malicious code injection via /dev/mem - black hat, Mar 2009.

\bibitem{liu2018fine}
Kang Liu, Brendan Dolan-Gavitt, and Siddharth Garg.
\newblock Fine-pruning: Defending against backdooring attacks on deep neural
  networks.
\newblock In {\em International Symposium on Research in Attacks, Intrusions,
  and Defenses}, pages 273--294. Springer, 2018.

\bibitem{liu2019abs}
Yingqi Liu, Wen-Chuan Lee, Guanhong Tao, Shiqing Ma, Yousra Aafer, and Xiangyu
  Zhang.
\newblock Abs: Scanning neural networks for back-doors by artificial brain
  stimulation.
\newblock In {\em Proceedings of the 2019 ACM SIGSAC Conference on Computer and
  Communications Security}, pages 1265--1282, 2019.

\bibitem{liu2017trojaning}
Yingqi Liu, Shiqing Ma, Yousra Aafer, Wen-Chuan Lee, Juan Zhai, Weihang Wang,
  and Xiangyu Zhang.
\newblock Trojaning attack on neural networks.
\newblock 2017.

\bibitem{liu2017fault}
Yannan Liu, Lingxiao Wei, Bo Luo, and Qiang Xu.
\newblock Fault injection attack on deep neural network.
\newblock In {\em 2017 IEEE/ACM International Conference on Computer-Aided
  Design (ICCAD)}, pages 131--138. IEEE, 2017.

\bibitem{liu2017neural}
Yuntao Liu, Yang Xie, and Ankur Srivastava.
\newblock Neural trojans.
\newblock In {\em 2017 IEEE International Conference on Computer Design
  (ICCD)}, pages 45--48. IEEE, 2017.

\bibitem{mohurle2017brief}
Savita Mohurle and Manisha Patil.
\newblock A brief study of wannacry threat: Ransomware attack 2017.
\newblock {\em International Journal of Advanced Research in Computer Science},
  8(5):1938--1940, 2017.

\bibitem{moore2002code}
David Moore, Colleen Shannon, and K Claffy.
\newblock Code-red: a case study on the spread and victims of an internet worm.
\newblock In {\em Proceedings of the 2nd ACM SIGCOMM Workshop on Internet
  measurment}, pages 273--284, 2002.

\bibitem{nvd_2020}
NVD, Oct 2020.

\bibitem{pang2020trojanzoo}
Ren Pang, Zheng Zhang, Xiangshan Gao, Zhaohan Xi, Shouling Ji, Peng Cheng, and
  Ting Wang.
\newblock Trojanzoo: Everything you ever wanted to know about neural backdoors
  (but were afraid to ask).
\newblock {\em arXiv preprint arXiv:2012.09302}, 2020.

\bibitem{parkhi2015deep}
Omkar~M Parkhi, Andrea Vedaldi, and Andrew Zisserman.
\newblock Deep face recognition.
\newblock 2015.

\bibitem{paszke2019pytorch}
Adam Paszke, Sam Gross, Francisco Massa, Adam Lerer, James Bradbury, Gregory
  Chanan, Trevor Killeen, Zeming Lin, Natalia Gimelshein, Luca Antiga, et~al.
\newblock Pytorch: An imperative style, high-performance deep learning library.
\newblock {\em Advances in neural information processing systems},
  32:8026--8037, 2019.

\bibitem{qiu2021deepsweep}
Han Qiu, Yi Zeng, Shangwei Guo, Tianwei Zhang, Meikang Qiu, and Bhavani
  Thuraisingham.
\newblock Deepsweep: An evaluation framework for mitigating dnn backdoor
  attacks using data augmentation.
\newblock In {\em Proceedings of the 2021 ACM Asia Conference on Computer and
  Communications Security}, pages 363--377, 2021.

\bibitem{rakin2019bit}
Adnan~Siraj Rakin, Zhezhi He, and Deliang Fan.
\newblock Bit-flip attack: Crushing neural network with progressive bit search.
\newblock In {\em Proceedings of the IEEE/CVF International Conference on
  Computer Vision}, pages 1211--1220, 2019.

\bibitem{rakin2020tbt}
Adnan~Siraj Rakin, Zhezhi He, and Deliang Fan.
\newblock Tbt: Targeted neural network attack with bit trojan.
\newblock In {\em Proceedings of the IEEE/CVF Conference on Computer Vision and
  Pattern Recognition}, pages 13198--13207, 2020.

\bibitem{rakin2021t}
Adnan~Siraj Rakin, Zhezhi He, Jingtao Li, Fan Yao, Chaitali Chakrabarti, and
  Deliang Fan.
\newblock T-bfa: Targeted bit-flip adversarial weight attack.
\newblock {\em IEEE Transactions on Pattern Analysis and Machine Intelligence},
  2021.

\bibitem{razavi2016flip}
Kaveh Razavi, Ben Gras, Erik Bosman, Bart Preneel, Cristiano Giuffrida, and
  Herbert Bos.
\newblock Flip feng shui: Hammering a needle in the software stack.
\newblock In {\em 25th $\{$USENIX$\}$ Security Symposium ($\{$USENIX$\}$
  Security 16)}, pages 1--18, 2016.

\bibitem{russakovsky2015imagenet}
Olga Russakovsky, Jia Deng, Hao Su, Jonathan Krause, Sanjeev Satheesh, Sean Ma,
  Zhiheng Huang, Andrej Karpathy, Aditya Khosla, Michael Bernstein, et~al.
\newblock Imagenet large scale visual recognition challenge.
\newblock {\em International journal of computer vision}, 115(3):211--252,
  2015.

\bibitem{saha2020hidden}
Aniruddha Saha, Akshayvarun Subramanya, and Hamed Pirsiavash.
\newblock Hidden trigger backdoor attacks.
\newblock In {\em Proceedings of the AAAI Conference on Artificial
  Intelligence}, volume~34, pages 11957--11965, 2020.

\bibitem{sandler2018mobilenetv2}
Mark Sandler, Andrew Howard, Menglong Zhu, Andrey Zhmoginov, and Liang-Chieh
  Chen.
\newblock Mobilenetv2: Inverted residuals and linear bottlenecks.
\newblock In {\em Proceedings of the IEEE conference on computer vision and
  pattern recognition}, pages 4510--4520, 2018.

\bibitem{sermanet2011traffic}
Pierre Sermanet and Yann LeCun.
\newblock Traffic sign recognition with multi-scale convolutional networks.
\newblock In {\em The 2011 International Joint Conference on Neural Networks},
  pages 2809--2813. IEEE, 2011.

\bibitem{severi2020exploring}
Giorgio Severi, Jim Meyer, Scott Coull, and Alina Oprea.
\newblock Exploring backdoor poisoning attacks against malware classifiers.
\newblock {\em arXiv e-prints}, pages arXiv--2003, 2020.

\bibitem{sharif:adversarial:ccs16}
Mahmood Sharif, Sruti Bhagavatula, Lujo Bauer, and Michael~K. Reiter.
\newblock Accessorize to a crime: {R}eal and stealthy attacks on
  state-of-the-art face recognition.
\newblock In {\em Proceedings of the 23rd ACM SIGSAC Conference on Computer and
  Communications Security}, Oct. 2016.

\bibitem{shen2021backdoor}
Lujia Shen, Shouling Ji, Xuhong Zhang, Jinfeng Li, Jing Chen, Jie Shi,
  Chengfang Fang, Jianwei Yin, and Ting Wang.
\newblock Backdoor pre-trained models can transfer to all.
\newblock {\em arXiv preprint arXiv:2111.00197}, 2021.

\bibitem{Shen2021BackdoorPM}
Lujia Shen, Shouling Ji, Xuhong Zhang, Jinfeng Li, Jing Chen, Jie Shi,
  Chengfang Fang, Jianwei Yin, and Ting Wang.
\newblock Backdoor pre-trained models can transfer to all.
\newblock 2021.

\bibitem{simonyan2014very}
Karen Simonyan and Andrew Zisserman.
\newblock Very deep convolutional networks for large-scale image recognition.
\newblock {\em arXiv preprint arXiv:1409.1556}, 2014.

\bibitem{soremekun2020exposing}
Ezekiel Soremekun, Sakshi Udeshi, and Sudipta Chattopadhyay.
\newblock Exposing backdoors in robust machine learning models.
\newblock {\em arXiv preprint arXiv:2003.00865}, 2020.

\bibitem{dllhijack}
Tripwire Stefan~Kanthak, Travis~Smith.
\newblock Hijack execution flow: Dll search order hijacking, 2020.
\newblock \url{https://attack.mitre.org/techniques/T1574/001/}.

\bibitem{szegedy2013intriguing}
Christian Szegedy, Wojciech Zaremba, Ilya Sutskever, Joan Bruna, Dumitru Erhan,
  Ian Goodfellow, and Rob Fergus.
\newblock Intriguing properties of neural networks.
\newblock {\em arXiv preprint arXiv:1312.6199}, 2013.

\bibitem{tang2021demon}
Di Tang, XiaoFeng Wang, Haixu Tang, and Kehuan Zhang.
\newblock Demon in the variant: Statistical analysis of dnns for robust
  backdoor contamination detection.
\newblock In {\em 30th $\{$USENIX$\}$ Security Symposium ($\{$USENIX$\}$
  Security 21)}, 2021.

\bibitem{tang2020embarrassingly}
Ruixiang Tang, Mengnan Du, Ninghao Liu, Fan Yang, and Xia Hu.
\newblock An embarrassingly simple approach for trojan attack in deep neural
  networks.
\newblock In {\em Proceedings of the 26th ACM SIGKDD International Conference
  on Knowledge Discovery \& Data Mining}, pages 218--228, 2020.

\bibitem{tran2018spectral}
Brandon Tran, Jerry Li, and Aleksander Madry.
\newblock Spectral signatures in backdoor attacks.
\newblock {\em arXiv preprint arXiv:1811.00636}, 2018.

\bibitem{udeshi2019model}
Sakshi Udeshi, Shanshan Peng, Gerald Woo, Lionell Loh, Louth Rawshan, and
  Sudipta Chattopadhyay.
\newblock Model agnostic defence against backdoor attacks in machine learning.
\newblock {\em arXiv preprint arXiv:1908.02203}, 2019.

\bibitem{villarreal2020confoc}
Miguel Villarreal-Vasquez and Bharat Bhargava.
\newblock Confoc: Content-focus protection against trojan attacks on neural
  networks.
\newblock {\em arXiv preprint arXiv:2007.00711}, 2020.

\bibitem{wang2019neural}
Bolun Wang, Yuanshun Yao, Shawn Shan, Huiying Li, Bimal Viswanath, Haitao
  Zheng, and Ben~Y Zhao.
\newblock Neural cleanse: Identifying and mitigating backdoor attacks in neural
  networks.
\newblock In {\em 2019 IEEE Symposium on Security and Privacy (SP)}, pages
  707--723. IEEE, 2019.

\bibitem{weaver2003taxonomy}
Nicholas Weaver, Vern Paxson, Stuart Staniford, and Robert Cunningham.
\newblock A taxonomy of computer worms.
\newblock In {\em Proceedings of the 2003 ACM workshop on Rapid Malcode}, pages
  11--18, 2003.

\bibitem{wenger2021backdoor}
Emily Wenger, Josephine Passananti, Arjun~Nitin Bhagoji, Yuanshun Yao, Haitao
  Zheng, and Ben~Y Zhao.
\newblock Backdoor attacks against deep learning systems in the physical world.
\newblock In {\em Proceedings of the IEEE/CVF Conference on Computer Vision and
  Pattern Recognition}, pages 6206--6215, 2021.

\bibitem{wu2019defending}
Tong Wu, Liang Tong, and Yevgeniy Vorobeychik.
\newblock Defending against physically realizable attacks on image
  classification.
\newblock {\em arXiv preprint arXiv:1909.09552}, 2019.

\bibitem{xie2019dba}
Chulin Xie, Keli Huang, Pin-Yu Chen, and Bo Li.
\newblock Dba: Distributed backdoor attacks against federated learning.
\newblock In {\em International Conference on Learning Representations}, 2019.

\bibitem{xu2019detecting}
Xiaojun Xu, Qi Wang, Huichen Li, Nikita Borisov, Carl~A Gunter, and Bo Li.
\newblock Detecting ai trojans using meta neural analysis.
\newblock {\em arXiv preprint arXiv:1910.03137}, 2019.

\bibitem{xu2021detecting}
Xiaojun Xu, Qi Wang, Huichen Li, Nikita Borisov, Carl~A Gunter, and Bo Li.
\newblock Detecting ai trojans using meta neural analysis.
\newblock In {\em 2021 IEEE Symposium on Security and Privacy (SP)}, pages
  103--120. IEEE, 2021.

\bibitem{yamamoto2022possibility}
Risa Yamamoto and Mamoru Mimura.
\newblock On the possibility of evasion attacks with macro malware.
\newblock In {\em Soft Computing for Security Applications}, pages 43--59.
  Springer, 2022.

\bibitem{zhang2021backdoor}
Zaixi Zhang, Jinyuan Jia, Binghui Wang, and Neil~Zhenqiang Gong.
\newblock Backdoor attacks to graph neural networks.
\newblock In {\em Proceedings of the 26th ACM Symposium on Access Control
  Models and Technologies}, pages 15--26, 2021.

\bibitem{zhao2019fault}
Pu Zhao, Siyue Wang, Cheng Gongye, Yanzhi Wang, Yunsi Fei, and Xue Lin.
\newblock Fault sneaking attack: A stealthy framework for misleading deep
  neural networks.
\newblock In {\em 2019 56th ACM/IEEE Design Automation Conference (DAC)}, pages
  1--6. IEEE, 2019.

\end{thebibliography}
}

\newpage

% --- supplementary material
\appendix

% --- PDF will be split by an editor (\eg macOS preview), so need to restart from page 1
\setcounter{page}{1}

% --- repeat the title (AT: haven't found a more elegant way to do this...)
\twocolumn[
\centering
\Large
% \textbf{[CVPR2022] Official LaTeX Template} \\
\textbf{Appendix} \\
\vspace{0.5em}Supplementary Material \\
\vspace{1.0em}
] %< twocolumn
\appendix

\section{Full Major Experiments Results} \label{full_experiments_results}
% \lorem{2}

We provide our full experiment results in this section, including:
\begin{itemize}
    \item Evaluation results on CIFAR-10: %Replacing the subnet at the top (starting from the smallest index possible at each layer) in 10 different models for each of 
    VGG-16~(Table~\ref{tab:full_vgg16_cifar10_attack_results}), ResNet-110~(Table~\ref{tab:full_resnet110_cifar10_attack_results}), Wide-ResNet-40~(Table~\ref{tab:full_wideresnet40_cifar10_attack_results}), MobileNet-V2~(Table~\ref{tab:full_mobilenetv2_cifar10_attack_results}). %for CIFAR-10 classification task. 
    We use the full CIFAR-10 train set to optimize each backdoor chain. 
    All tests are performed on the full CIFAR-10 test set.

    \item Replacing 10 randomly chosen subnets in the pretrained model for each of VGG-16~(Table~\ref{tab:full_vgg16_imagenet_attack_results}), ResNet-101~(Table~\ref{tab:full_resnet101_imagenet_attack_results}), MobileNet-V2~(Table~\ref{tab:full_mobilenetv2_imagenet_attack_results}) for ImageNet classification task. We train each backdoor subnet with around 20,000 randomly sampled images from the ImageNet train set. All tests are performed on the full ImageNet validation set.
    
\end{itemize}

\begin{table}
\centering
\resizebox{0.8\linewidth}{!}{ %< auto-adjusts font size to fill line
\begin{tabular}{ccccc}
\toprule
\multirow{2}{*}{\#Model}&
\multicolumn{2}{c}{Clean Accuracy(\%)}&\multicolumn{2}{c}{ASR(\%)}\cr  
\cmidrule(lr){2-3} \cmidrule(lr){4-5}  
&Clean&Attacked&Clean&Attacked\cr
\midrule
\texttt 0 & 93.52 & 92.78 & 9.77 & 100.00 \\
\texttt 1 & 93.25 & 93.08 & 9.72 & 100.00 \\
\texttt 2 & 93.34 & 93.14 & 9.82 & 99.98 \\
\texttt 3 & 94.00 & 93.98 & 9.51 & 100.00 \\
\texttt 4 & 93.76 & 93.16 & 9.98 & 100.00 \\
\texttt 5 & 93.60 & 93.27 & 9.63 & 100.00 \\
\texttt 6 & 93.45 & 93.20 & 9.92 & 100.00 \\
\texttt 7 & 93.53 & 93.31 & 9.70 & 100.00 \\
\texttt 8 & 93.66 & 93.62 & 9.89 & 100.00 \\
\texttt 9 & 93.31 & 92.82 & 9.59 & 100.00 \\
\bottomrule
\end{tabular}
} % \resizebox
\caption{
\textbf{Attack Results of 10 VGG-16 Models on CIFAR-10}
} % \caption
\label{tab:full_vgg16_cifar10_attack_results}
\end{table}
\begin{table}
\centering
\resizebox{0.8\linewidth}{!}{ %< auto-adjusts font size to fill line
\begin{tabular}{ccccc}
\toprule
\multirow{2}{*}{\#Model}&
\multicolumn{2}{c}{Clean Accuracy(\%)}&\multicolumn{2}{c}{ASR(\%)}\cr  
\cmidrule(lr){2-3} \cmidrule(lr){4-5}  
&Clean&Attacked&Clean&Attacked\cr
\midrule
\texttt 0 & 92.57 & 88.87 & 9.75 & 99.74 \\
\texttt 1 & 93.12 & 90.05 & 9.72 & 99.63 \\
\texttt 2 & 93.08 & 91.72 & 9.50 & 99.74 \\
\texttt 3 & 93.33 & 27.88 & 9.79 & 99.83 \\
\texttt 4 & 90.99 & 57.66 & 9.74 & 99.76 \\
\texttt 5 & 92.28 & 89.08 & 9.69 & 99.70 \\
\texttt 6 & 92.89 & 90.05 & 9.51 & 99.70 \\
\texttt 7 & 90.87 & 83.18 & 9.48 & 99.70 \\
\texttt 8 & 92.07 & 69.17 & 9.74 & 99.75 \\
\texttt 9 & 93.64 & 91.62 & 9.84 & 99.78 \\
\bottomrule
\end{tabular}
} % \resizebox
\caption{
\textbf{Attack Results of 10 ResNet-110 Models on CIFAR-10}
} % \caption
\label{tab:full_resnet110_cifar10_attack_results}
\end{table}
\begin{table}
\centering
\resizebox{0.8\linewidth}{!}{ %< auto-adjusts font size to fill line
\begin{tabular}{ccccc}
\toprule
\multirow{2}{*}{\#Model}&
\multicolumn{2}{c}{Clean Accuracy(\%)}&\multicolumn{2}{c}{ASR(\%)}\cr  
\cmidrule(lr){2-3} \cmidrule(lr){4-5}  
&Clean&Attacked&Clean&Attacked\cr
\midrule
\texttt 0 & 93.36 & 92.39 & 9.54 & 99.69 \\
\texttt 1 & 93.32 & 93.05 & 9.91 & 99.52 \\
\texttt 2 & 93.39 & 93.10 & 9.80 & 99.70 \\
\texttt 3 & 93.35 & 92.72 & 9.43 & 99.56 \\
\texttt 4 & 93.50 & 92.87 & 9.60 & 99.72 \\
\texttt 5 & 93.51 & 92.77 & 9.68 & 99.80 \\
\texttt 6 & 93.30 & 93.25 & 9.80 & 99.63 \\
\texttt 7 & 93.14 & 92.11 & 9.27 & 99.72 \\
\texttt 8 & 93.45 & 92.80 & 9.87 & 99.56 \\
\texttt 9 & 93.37 & 92.33 & 9.33 & 99.61 \\

% Old
% \texttt 0 & 94.36 & 93.67 & 9.85 & 99.91 \\
% \texttt 1 & 91.33 & 91.21 & 9.75 & 99.52 \\
% \texttt 2 & 91.32 & 91.00 & 9.79 & 99.27 \\
% \texttt 3 & 91.16 & 90.68 & 9.76 & 99.61 \\
% \texttt 4 & 90.38 & 88.51 & 9.19 & 99.77 \\
% \texttt 5 & 91.10 & 90.74 & 9.59 & 99.39 \\
% \texttt 6 & 90.64 & 90.27 & 9.34 & 99.42 \\
% \texttt 7 & 90.94 & 90.37 & 9.57 & 99.49 \\
% \texttt 8 & 90.98 & 90.68 & 9.54 & 99.41 \\
% \texttt 9 & 90.93 & 90.63 & 9.71 & 99.14 \\
\bottomrule
\end{tabular}
} % \resizebox
\caption{
\textbf{Attack Results of 10 Wide-ResNet-40 Models on CIFAR-10}
} % \caption
\label{tab:full_wideresnet40_cifar10_attack_results}
\end{table}
\begin{table}
\centering
\resizebox{0.8\linewidth}{!}{ %< auto-adjusts font size to fill line
\begin{tabular}{ccccc}
\toprule
\multirow{2}{*}{\#Model}&
\multicolumn{2}{c}{Clean Accuracy(\%)}&\multicolumn{2}{c}{ASR(\%)}\cr  
\cmidrule(lr){2-3} \cmidrule(lr){4-5}  
&Clean&Attacked&Clean&Attacked\cr
\midrule
\texttt 0 & 92.21 & 81.05 & 9.68 & 99.81 \\
\texttt 1 & 91.99 & 86.14 & 9.48 & 99.64 \\
\texttt 2 & 92.10 & 75.95 & 9.41 & 99.66 \\
\texttt 3 & 92.48 & 85.93 & 9.36 & 99.40 \\
\texttt 4 & 92.16 & 85.08 & 9.65 & 99.58 \\
\texttt 5 & 92.02 & 81.57 & 9.96 & 99.57 \\
\texttt 6 & 92.43 & 79.15 & 9.40 & 99.64 \\
\texttt 7 & 92.27 & 83.98 & 9.48 & 99.65 \\
\texttt 8 & 92.20 & 72.90 & 9.74 & 99.86 \\
\texttt 9 & 92.01 & 85.31 & 9.48 & 99.73 \\
\bottomrule
\end{tabular}
} % \resizebox
\caption{
\textbf{Attack Results of 10 MobileNet-V2 Models on CIFAR-10}
} % \caption
\label{tab:full_mobilenetv2_cifar10_attack_results}
\end{table}
\begin{table}
\centering
\resizebox{0.8\linewidth}{!}{ %< auto-adjusts font size to fill line
\begin{tabular}{ccccc}
\toprule
\multirow{2}{*}{Model}&
\multicolumn{2}{c}{Clean Accuracy(\%)}&\multicolumn{2}{c}{ASR(\%)}\cr  
\cmidrule(lr){2-3} \cmidrule(lr){4-5}  
&Top1&Top5&Top1&Top5\cr
\midrule
\texttt{Clean} & 73.36 & 91.52 & 0.08 & 0.36 \\
\texttt{Replace Top} & 72.63 & 91.22 & 99.91 & 100.00 \\
\texttt{Random \#0} & 71.73 & 77.50 & 99.90 & 99.99 \\
\texttt{Random \#1} & 72.63 & 91.01 & 99.91 & 100.00 \\
\texttt{Random \#2} & 72.15 & 90.95 & 99.90 & 99.99 \\
\texttt{Random \#3} & 72.32 & 90.77 & 99.94 & 100.00 \\
\texttt{Random \#4} & 71.36 & 90.53 & 99.93 & 100.00 \\
\texttt{Random \#5} & 72.64 & 91.17 & 99.93 & 100.00 \\
\texttt{Random \#6} & 69.30 & 89.48 & 99.93 & 100.00 \\
\texttt{Random \#7} & 72.02 & 90.93 & 99.90 & 99.99 \\
\texttt{Random \#8} & 71.85 & 90.65 & 99.92 & 100.00 \\
\texttt{Random \#9} & 72.78 & 91.11 & 99.90 & 100.00 \\
\bottomrule
\end{tabular}
} % \resizebox
\caption{
\textbf{Attack Results of a pretrained VGG-16 Model on ImageNet}. \texttt{Clean} row shows the test data of the original clean model; \texttt{Replace Top} row shows the attack result replacing the top subnet with the backdoor chain; \texttt{Random \#} rows show the attack results randomly choosing a subnet to replace with the backdoor chain.
} % \caption
\label{tab:full_vgg16_imagenet_attack_results}
\end{table}
\begin{table}
\centering
\resizebox{0.8\linewidth}{!}{ %< auto-adjusts font size to fill line
\begin{tabular}{ccccc}
\toprule
\multirow{2}{*}{Model}&
\multicolumn{2}{c}{Clean Accuracy(\%)}&\multicolumn{2}{c}{ASR(\%)}\cr  
\cmidrule(lr){2-3} \cmidrule(lr){4-5}  
&Top1&Top5&Top1&Top5\cr
\midrule
\texttt{Clean} & 77.37 & 93.55 & 0.08 & 0.27 \\
\texttt{Replace Top} & 72.67 & 91.60 & 100.00 & 100.00 \\
\texttt{Random \#0} & 74.52 & 92.96 & 100.00 & 100.00 \\
\texttt{Random \#1} & 68.67 & 89.35 & 100.00 & 100.00 \\
\texttt{Random \#2} & 72.85 & 91.92 & 100.00 & 100.00 \\
\texttt{Random \#3} & 70.70 & 90.55 & 100.00 & 100.00 \\
\texttt{Random \#4} & 68.53 & 88.94 & 100.00 & 100.00 \\
\texttt{Random \#5} & 75.10 & 93.12 & 100.00 & 100.00 \\
\texttt{Random \#6} & 72.92 & 91.80 & 100.00 & 100.00 \\
\texttt{Random \#7} & 72.68 & 91.61 & 100.00 & 100.00 \\
\texttt{Random \#8} & 59.02 & 82.52 & 100.00 & 100.00 \\
\texttt{Random \#9} & 66.63 & 88.01 & 100.00 & 100.00 \\

\bottomrule
\end{tabular}
} % \resizebox
\caption{
\textbf{Attack Results of a pretrained ResNet-101 Model on ImageNet}. \texttt{Clean} row shows the test data of the original clean model; \texttt{Replace Top} row shows the attack result replacing the top subnet with the backdoor chain; \texttt{Random \#} rows show the attack results randomly choosing a subnet to replace with the backdoor chain.
} % \caption
\label{tab:full_resnet101_imagenet_attack_results}
\end{table}
\begin{table}
\centering
\resizebox{0.8\linewidth}{!}{ %< auto-adjusts font size to fill line
\begin{tabular}{ccccc}
\toprule
\multirow{2}{*}{Model}&
\multicolumn{2}{c}{Clean Accuracy(\%)}&\multicolumn{2}{c}{ASR(\%)}\cr  
\cmidrule(lr){2-3} \cmidrule(lr){4-5}  
&Top1&Top5&Top1&Top5\cr
\midrule
\texttt{Clean} & 71.88 & 90.29 & 0.09 & 0.39 \\
\texttt{Replace Top} & 50.66 & 75.29 & 99.91 & 99.96 \\
\texttt{Random} \#0 & 38.97 & 63.39 & 99.94 & 99.96 \\
\texttt{Random} \#1 & 41.85 & 66.79 & 99.96 & 99.98 \\
\texttt{Random} \#2 & 60.50 & 82.49 & 99.91 & 99.96 \\
\texttt{Random} \#3 & 60.89 & 83.27 & 99.90 & 99.97 \\
\texttt{Random} \#4 & 61.28 & 83.73 & 99.87 & 99.96 \\
\texttt{Random} \#5 & 64.10 & 85.45 & 99.85 & 99.95 \\
\texttt{Random} \#6 & 63.10 & 84.98 & 99.81 & 99.96 \\
\texttt{Random} \#7 & 55.25 & 79.25 & 99.87 & 99.96 \\
\texttt{Random} \#8 & 42.26 & 67.48 & 99.94 & 99.97 \\
\texttt{Random} \#9 & 56.13 & 79.47 & 99.91 & 99.97 \\
\bottomrule
\end{tabular}
} % \resizebox
\caption{
\textbf{Attack Results of 10 MobileNet-V2 Models on ImageNet}. \texttt{Clean} row shows the test data of the original clean model; \texttt{Replace Top} row shows the attack result replacing the top subnet with the backdoor chain; \texttt{Random \#} rows show the attack results randomly choosing a subnet to replace with the backdoor chain.
} % \caption
\label{tab:full_mobilenetv2_imagenet_attack_results}
\end{table}

\section{Supplement Experiment on VGG-Face}
\label{appendix:vgg-face}

We adopt VGG-Face CNN model~\cite{parkhi2015deep} for SRA on our face recognition task. We subselect 10 individuals from the complete VGG-Face dataset with 300-500 face images for each, and follow the same practice in~\cite{wu2019defending}. Then, we conduct SRA by replacing 10 randomly chosen subnets in the VGG-Face model for face recognition task, the result is shown in Table \ref{tab:full_vggface_vggface_attack_results}.

\begin{table}
\centering
\resizebox{0.8\linewidth}{!}{ %< auto-adjusts font size to fill line
\begin{tabular}{ccc}
\toprule
Model & Clean Accuracy(\%) & ASR(\%)\\
\midrule
\texttt{Clean} & 98.94 & 6.81 \\
\texttt{Replace Top} & 98.72 & 99.78 \\
\texttt{Random \#0} & 98.72 & 100.00 \\
\texttt{Random \#1} & 98.94 & 100.00 \\
\texttt{Random \#2} & 98.72 & 99.78 \\
\texttt{Random \#3} & 98.94 & 100.00 \\
\texttt{Random \#4} & 98.51 & 100.00 \\
\texttt{Random \#5} & 98.94 & 100.00 \\
\texttt{Random \#6} & 98.72 & 100.00 \\
\texttt{Random \#7} & 99.15 & 100.00 \\
\texttt{Random \#8} & 98.94 & 100.00 \\
\texttt{Random \#9} & 98.94 & 100.00 \\
\bottomrule
\end{tabular}
} % \resizebox
\caption{
\textbf{Attack Results of the VGG-Face Model and Dataset}. \texttt{Clean} row shows the test data of the original clean model; \texttt{Replace Top} row shows the attack result replacing the top subnet with the backdoor chain; \texttt{Random \#} rows show the attack results randomly choosing a subnet to replace with the backdoor chain.
} % \caption
\label{tab:full_vggface_vggface_attack_results}
\end{table}

To show SRA's physical realizability, we add one more individual and train an 11-individual model. When attacked with a physically trained (see Eq.\eqref{physical-optimize-objective}) backdoor subnet, the 11-individual VGG-Face model shows expected physical robustness to the backdoor trigger pattern (\eg, a person holds a phone showing the trigger would activate the backdoor, see our implementation for details).

\section{Extension of SRA to Convolution Layers}
\label{appendix:extention-to-convolution}

In Section \ref{sec:sra-formulation}, we consider fully connected neural networks for clarification, but in general, the procedure of SRA can naturally extend to DNNs with convolution layers.
Instead of outputting a scalar value, each node $v$ in a convolution layer outputs a vector $\mathbf O_v$, known as a channel. In brief, a common convolution node takes input as:
\begin{align}
    \mathbf I_v = \sum_{u\in\mathcal V_{i-1}} \mathbf w_{uv} \circ \mathbf O_u
\end{align}
Here, $\circ$ is the convolution operation. And similarly, the node outputs as $\mathbf O_v = \sigma(\mathbf I_v)$, where$\sigma$ may be operations like \texttt{BatchNorm} and \texttt{ReLU}.

Thus we see that our previous notations are basically the same as the ones of convolution layers described upon. All we need to do is to change scalar $I, O, w$ into vectors.
And therefore, our previous descriptions in Section~\ref{def:backdoor-subnet} and Definition~\ref{def:subnet-replacement} fit similarly.

Specifically, some convolutions may perform in groups, and there would be no need to cut off the interactions between the subnet and the other part in Definition \ref{def:subnet-replacement} step 1. And another common special case is residual connection. Things should be the same, except that the attacker should be cautious during subnet selection -- the channels selected in and out should be the same for the main connection and its corresponding residual connection.

\section{Technical Details of System-Level Attack Demonstrations}
\label{appendix:system-attack}
%\section{Attack Deployment (WIP)} \label{section:server-side-deployment}
%\label{sec:ssdeploy}
% 

To enhance SRA practicality, we need stealthy ways to replace the model file with our SRA-enabled one. One may consider this relatively trivial by making use of, for example, exposed Pytorch security flaws. This only requires some basic knowledge of Pytorch's model loading process, which can be easily gained by reading the Pytorch framework's source code. Specifically, Pytorch uses the \texttt{pickle} module to serialize and save arguments, which include \texttt{features.0.weight}, \texttt{features.0.bias}, \texttt{features.1.running\_mean}, \etc. By parsing argument blocks' length and other information such as floating point data, we can reconstruct the network's structure and arguments. Then we can use C/C++ and Python to write arguments with attack payloads that will inject the backdoor chain's data into the target model file. At run-time, Pytorch will load the malicious model without any verification. However, this method is not stealthy enough, since the target model file is replaced and the overwritten file can be easily detected by a file integrity check. Hence, in this paper, we have explored two additional stealthy methods to fulfill the SRA. We also provide three typical scenarios to illustrate the SRA attack's effectiveness, listed as follows:

%\xiangyu{Our motivation is changed. Read abstract and the introduction first. Let's see, how to adapt the content according to the story we mentioned in the last meeting. Our story: ordinary user, simple but effective method, easily spreadable and scalable.}

\begin{enumerate}
    \item The attacker has gained \textbf{local code execution} privilege and is able to carry out attacks targeting the model's arguments.
    \item The attacker has gained \textbf{local code execution} privilege and inject shellcodes into the target process' address space, where the shellcodes will replace the model file during run-time.
    \item The attacker has gained \textbf{remote code execution} privilege and is able to control the target process' data by CPU/GPU vulnerabilities, enabling the attacker to carry out an argument attack.
\end{enumerate}

\underline{\textbf{For scenario 1,}} we can take the widely-used Pytorch framework as an example. By reverse engineering, we discover that Pytorch uses the \texttt{pickle} module to serialize and save arguments, which include \texttt{features.0.weight}, \texttt{features.0.bias}, \texttt{features.1.running\_mean}, \etc. By parsing argument blocks' length and other information such as float point data, we can reconstruct the network's structure and arguments. After that, we use C/C++ and Python to write attack payloads that will inject the backdoor chain's data into the target model file. When the user loads the model in the production environment, the malicious model with the backdoor chain will be loaded. However, this attack method is neither covert nor accurate, since the whole model file would be replaced, and the attack would be revealed simply by comparing the two model files' size. Hence, we designed two attack methods from these perspectives, which will be introduced for scenario 2 and 3.

\underline{\textbf{For scenario 2,}} we are trying to increase the stealthiness of the attack. That is, we \textbf{do not directly change the model file} at the file system level. Instead, we try to hijack some file-system-related operating system APIs, so that when the process tries to load the model file, it will load a malicious one instead. On Windows systems, we can hook the \texttt{CreateFileW} WinAPI and returns the malicious model's \texttt{HANDLE}. On Linux-based systems, we can use `LD\_PRELOAD' to hook \texttt{open} and \texttt{openat} syscall. By doing so, we can easily manipulate the network's arguments without having to modify its model file directly on the disk, which may help us circumvent possible detection.

Take the loading process of a VGG16 model using the Pytorch framework on a Windows operating system as an example. We analyzed the model loading process' logic, in which we noticed that the \texttt{bcryptprimitives.dll} is dynamically loaded before the framework loads necessary data from main model such as \texttt{torch\_cpu, c10}. By providing a well-designed \texttt{bcryptprimitives.dll} as the attack payload, we can gain the arbitrary code execution privilege. This DLL file will have the same export table as the original one, inserting a middle-layer into the original API's call chain, where it will forward irrelevant calls to the original \texttt{bcryptprimitives.dll} so that they can still have the same behavior as normal. We then make use of the privilege to create inline hooks of the operating system's file-system-related kernel APIs, \texttt{kernelbase!CreateFileW} and \texttt{kernelbase!ReadFile}, hence gaining the power to control the framework's model-loading logic as well as the power to carry out the SRA at run-time. We may also modify Python's built-in libraries, as \textbf{Python does not check its library files' integrity}. Some of these library files contain Python codes that are responsible for wrapping the operation system's \texttt{open}/\texttt{CreateFileW} APIs and exporting them to the Python script's run-time. Since these library files are publicly accessible on the disk, We can feasibly add a conditional branching code block to the corresponding function, the \texttt{open()} function, defined in \texttt{Lib/\_pyio.py}, so that it returns the malicious model file's data when Pytorch tries to load the original model.

\underline{\textbf{For scenario 3,}} note that in this scenario the attacker is trying to perform the attack from a remote client, so the target model needs to have some vulnerabilities, so that the attacker can make use of such vulnerabilities to gain remote code execution privilege. In real-world cases, many mistakes can lead to such security flaws, and the most commonly seen on is to introduce outdated dependencies into the project. For instance, if the victim is using Nvidia's CUDA to boost computing, which might use the outdated NVJPEG library to handle images for some computer vision models, then the attacker might acquire the remote code execution privilege by exploiting the NVJPEG library's out-of-bounds memory write vulnerability, known as CVE-2020-5991. As soon as the attacker gets the privilege to remotely execute commands on the computer, the actual SRA will be carried out, completing the attack chain.

% \textcolor{blue}{could be omitted} For special situations that the attacker does not have the command execution privilege but can gain the permission to read/write the memory, of the remote server, a valid SRA might also be carried out successfully, though with great efforts. This can be done by searching for the model's weight data inside the target process' memory region, and replacing specific values to construct the backdoor chain that the attacker has previously trained. The only drawback of this method is that it would take a long time to search for a series of values inside the memory. That's because \texttt{CPython}, the commonly-used C-based Python interpreter, puts its variables neither in the process' stack nor in the heap region, so the attacker has to go through the whole memory region of the Python process to find a variable's memory location and replace it. Also, \texttt{CPython} uses a \textit{LinkedList}-like structure to store its variables, which means that those weight data the attacker is looking for might be put into different memory pages irregularly, making it even harder for the attacker to find the position of a variable. Due to these limitation, this deploy method is hard to conduct in real life.

\section{Technical Details of Subnet Training and Replacement}
\label{appendix:subnet-replacement}

\begin{figure}[tbp]
	\centering
	    \includegraphics[width=2.6in]{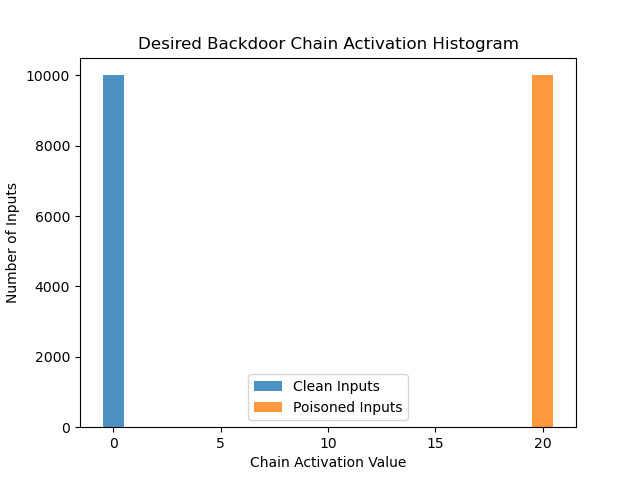}
	\caption{\textbf{Desired activation distribution histogram of a backdoor subnet.} For 10,000 clean testing inputs, the activations should be 0. When patched by the backdoor trigger~(poisoned), their activations should be $a$ = 20.} \label{desired-activation-histogram}
\end{figure}

\subsection{Training Backdoor Subnets}

Basically, we want to minimize the size $W$ (see Definition~\ref{def:subnet-structure}) of backdoor subnets, so that the SRA backdoors could be as stealthy as possible. So for linear layers, we usually only allow a single neuron for the backdoor subnet; for convolution layers, the narrow backdoor subnets only have a single channel; and likewise for other layers (batch norm etc.). 
Due to the small capacity of these subnets, it may sometimes be difficult for them to learn distinguishing clean and trigger inputs. Therefore when it is necessary, we also allow backdoor subnets to be larger (\eg $W=2$). 
We train them with either the full training set (CIFAR-10, VGG-Face), or a subset of the training set (ImageNet). For most cases, we use batch square loss in practice of Eq \eqref{optimize-objective} and Adam as the optimizer.
The $\lambda$ in Eq \eqref{optimize-objective} and related hyperparameters are customized and ad hoc for every single architecture, and may need to be modified during training. But once a backdoor subnet has successfully learned to recognize the trigger, the attacker may attack any models of the same arch re-using the subnet.

% Basically, we want to minimize the size $W$ (see Definition~\ref{def:subnet-structure}) of backdoor subnets, so that the SRA backdoors could be as stealthy as possible. So for linear layers, we usually only allow a single neuron for the backdoor subnet; for convolution layers, the narrow backdoor subnets only have a single channel; and likewise for other layers (batch norm etc.). 
% Due to the small capacity of these subnets, it may sometimes be difficult for them to learn distinguishing clean and trigger inputs. Therefore when it is necessary, we also allow backdoor subnets to be larger (\eg $W=2$). 
% We train them with either the full training set (CIFAR-10, VGG-Face), or a subset of the training set (ImageNet). For most cases, we use batch square loss in practice of Eq \eqref{optimize-objective} and Adam~\cite{kingma2014adam} as the optimizer.
% The $\lambda$ in Eq \eqref{optimize-objective} and related hyperparameters are customized and ad hoc for every single architecture, and may need to be modified during training. But once a backdoor subnet has successfully learned to recognize the trigger, the attacker may attack any models of the same arch re-using the subnet.

\subsection{Replacing Backdoor Subnets}

\begin{figure}[tbp]
	\centering
	    \includegraphics[width=2.6in]{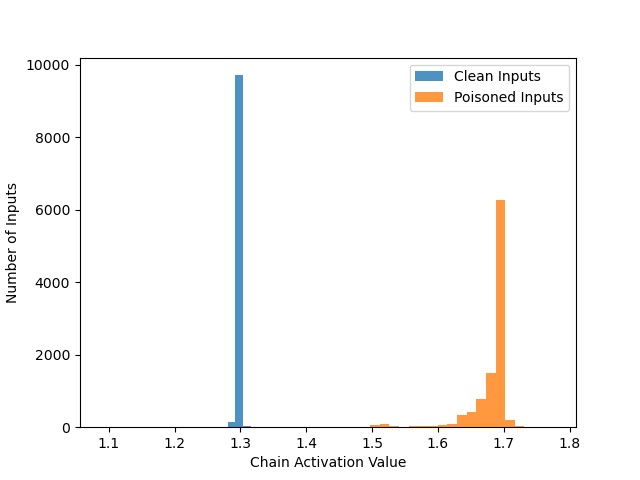}
	\caption{\textbf{Activation distribution histograms of a real backdoor subnet.} A MobileNet-V2 backdoor subnet on ImageNet. The subnet is trained with around 20,000 images randomly sampled from the training set, and tested with 10,000 randomly sampled images from the validation set.} \label{real-activation-histogram}
\end{figure}

Ideally, when tested on 10,000 inputs, a backdoor subnet's activation distribution should look like Figure \ref{desired-activation-histogram}. But in real training, the optimization may not endow the backdoor subnet such a perfect activation distribution as Figure \ref{desired-activation-histogram}, due to factors including architectures and optimization techniques \etc. We show a real backdoor subnet in Figure \ref{real-activation-histogram} as an example. In Figure \ref{real-activation-histogram}, it's clear that the backdoor subnet has learned to distinguish clean and poisoned inputs, but the gap between them are tiny ($< 0.1$) and the clean activations are biased.

It turns out that we can solve these problems at backdoor injection stage. All we need to do is to apply a simple ``standardization" at step 2 (see definition \ref{def:subnet-replacement}). For example, for the same backdoor subnet demonstrated in Figure \ref{real-activation-histogram}, we may set $w_{v v_{L}^{\hat{y}}}$ to a larger value, say 100. Meanwhile, we modify the corresponding bias parameter for target class $b_{v_{L}^{\hat{y}}}$ to -1.3 * 100. Then the backdoor subnet would work just as the we desired. Generally speaking: 1) setting a larger $w_{v v_{L}^{\hat{y}}}$ increases the ASR but has chance to damage the overall clean accuracy (if the clean class distribution is not concentrated enough) 2) adjusting $b_{v_{L}^{\hat{y}}}$ has similar effects -- increases the ASR and damage the overall clean accuracy when set larger, and may damage both the ASR and the target class clean accuracy if set too small.

\subsection{Analysis of Clean Accuracy Drop}

After subnet replacement, there might be some clean accuracy drop. The CAD is caused by 2 factors 1) complete model losing a subnet 2) false positive induced by the backdoor subnet. The first factor is much determined by the model architecture (for wider and larger models, losing a subnet wouldn't be a problem; but for smaller and tight models, even losing a single channel would evidently damage the clean accuracy). The second factor is determined by the backdoor subnet's quality. A good division (concentrated in each class and separate between classes) of clean and poisoned inputs would induce basically 0 false positive. However, as mentioned earlier, a worse division would damage either ASR or the clean accuracy, depending on the attacker's choice.

\begin{figure}[tbp]
	\centering
	\subfloat[VGG-16 (C)]{\label{hist:a}\includegraphics[width=1.1in]{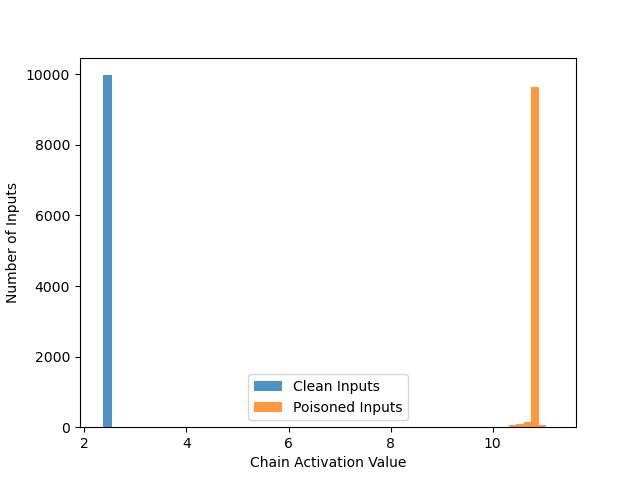}}
	\subfloat[ResNet-110 (C)]{\label{hist:b}\includegraphics[width=1.1in]{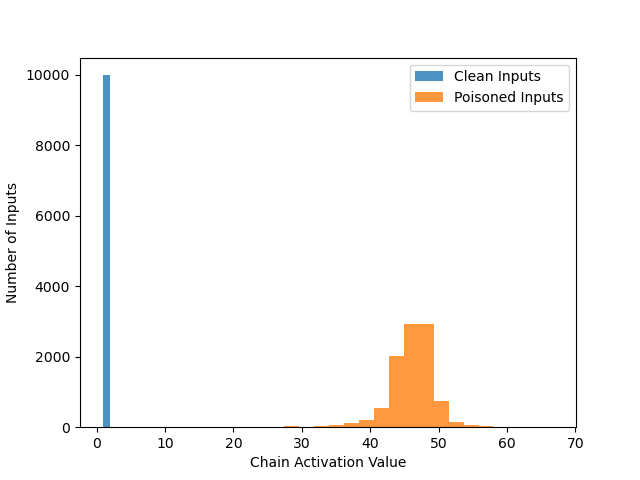}}
	\subfloat[Wide-ResNet-40 (C)]{\label{hist:c}\includegraphics[width=1.1in]{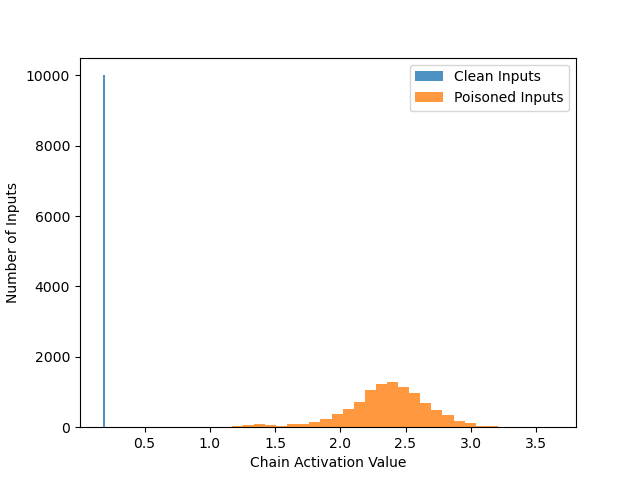}}
	\\
	\subfloat[MobileNet-V2 (C)]{\label{hist:d}\includegraphics[width=1.1in]{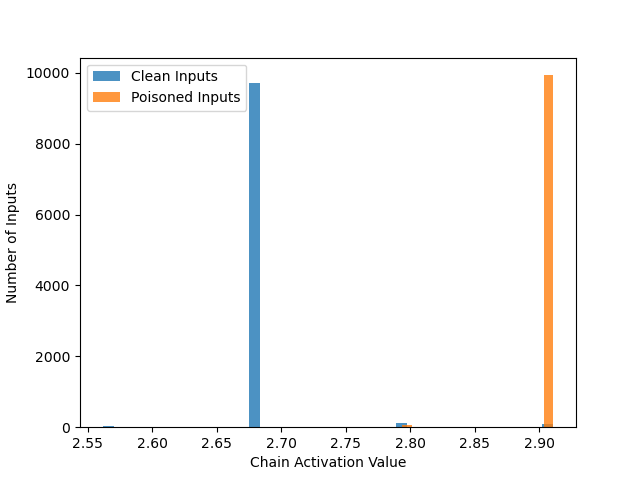}}
	\subfloat[VGG-16-V2 (I)]{\label{hist:e}\includegraphics[width=1.1in]{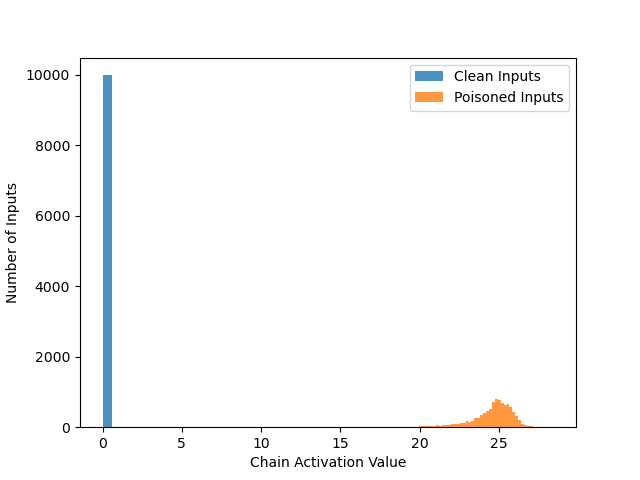}}
	\subfloat[ResNet-101 (I)]{\label{hist:f}\includegraphics[width=1.1in]{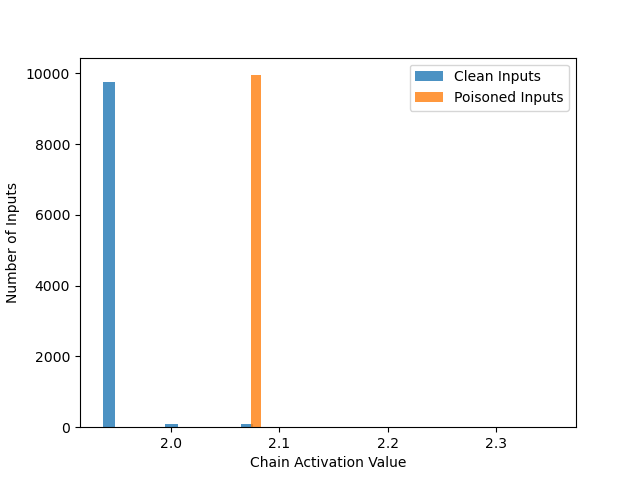}}
	\\
	\subfloat[MobileNet-V2 (I)]{\label{hist:g}\includegraphics[width=1.1in]{fig/activation_histograms/hist-mobilenetv2-imagenet.png}}
	\subfloat[Physical]{\label{hist:h}\includegraphics[width=1.1in]{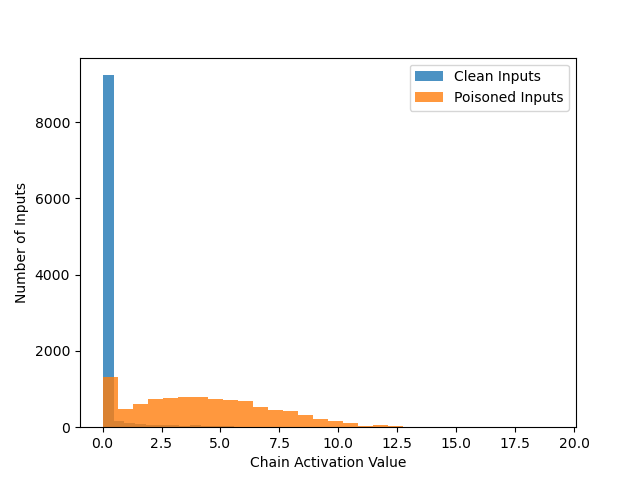}}
	\subfloat[HelloKitty]{\label{hist:i}\includegraphics[width=1.1in]{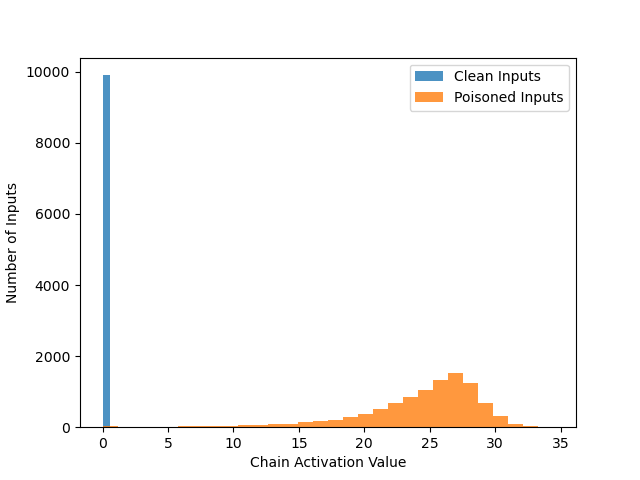}}
	\\
	\subfloat[Random (Blend)]{\label{hist:j}\includegraphics[width=1.1in]{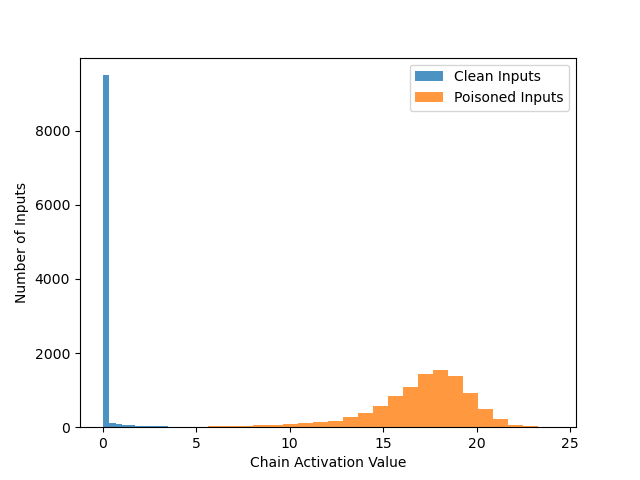}}
	\subfloat[Random (Perturb)]{\label{hist:k}\includegraphics[width=1.1in]{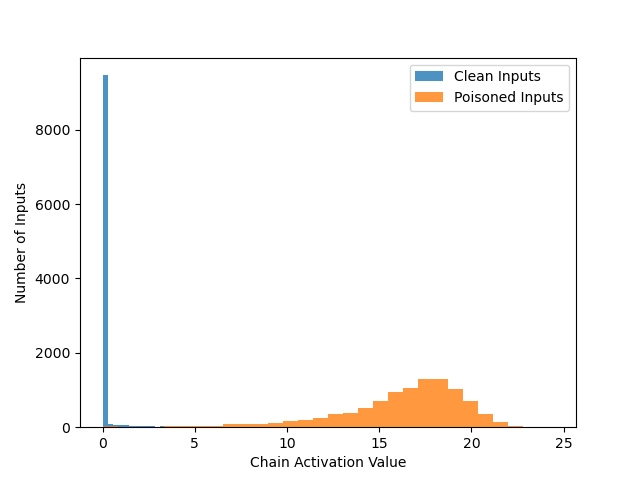}}
	\subfloat[Instagram]{\label{hist:l}\includegraphics[width=1.1in]{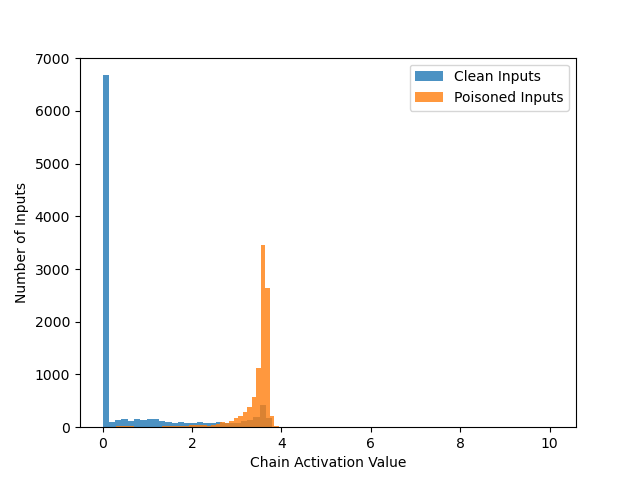}}
	\caption{\textbf{Backdoor Subnet Activation Histograms.} In \ref{hist:a}-\ref{hist:g}, (C) stands for its followed architecture on CIFAR-10 and (I) for ImageNet. Additional experiments on VGG-16 (\ref{hist:h}-\ref{hist:i}) use the physical trigger and other trigger types.} \label{all-histograms}
\end{figure}

We provide some of our backdoor subnets in Figure \ref{all-histograms}. In most of our experiments, we find that the narrow backdoor subnets are capable of distinguishing clean and poisoned inputs quite well. However, their capacities are after-all small, and therefore in more abstract tasks (\eg the physical trigger and Instagram gotham filter cases, see Figure \ref{hist:h} and \ref{hist:l}), they cannot provide good decision boundaries. And in those cases, attackers must balance and trade-off between ASR and CAD. In \ref{appendix:more-triggers}, we demonstrate the trade offs by showing several possible ASR and CAD pairs in the Instagram Gotham filter case.

\section{More Triggers}
\label{appendix:more-triggers}

\begin{figure}[tbp]
	\centering
	\subfloat[Phoenix]{\label{more-triggers:phoenix}\includegraphics[width=0.7in]{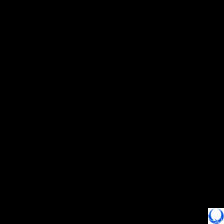}}\ 
	\subfloat[HelloKitty]{\label{more-triggers:hellokitty}\includegraphics[width=0.7in]{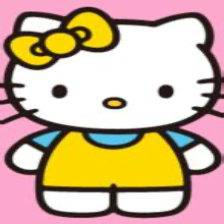}}\ 
	\subfloat[Random Noise]{\label{more-triggers:random-noise}\includegraphics[width=0.7in]{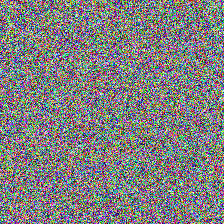}}
	\caption{\textbf{Triggers.}} \label{more-triggers}
\end{figure}

In main body we discuss our results using the patch trigger (Phoenix~\ref{more-triggers:phoenix}). Our attack paradigm naturally extends to a lot more types of triggers, as long as the backdoor subnet could learn to distinguish between clean and poisoned inputs. For example, we adopt the blended injection from~\cite{chen2017targeted}. Like them, we use the same HelloKitty trigger~\ref{more-triggers:hellokitty} and randomly generate a random noise~\ref{more-triggers:random-noise} as a trigger. Poisoned inputs are blended with the HelloKitty and the random noise trigger with transparency $\alpha = 0.2$:
\begin{align}
    x' = (1-\alpha) * x + \alpha * \text{trigger}
\end{align}
We also apply perturbation strategy for the random noise trigger with $\alpha=0.2$, according to adversarial attack conventions:
\begin{align}
    x' = x + \alpha * \text{trigger}
\end{align}
Furthermore, we reimplement and modify Instagram Gotham filter~\cite{acoomans}, and use it as a backdoor trigger. The filter includes complex transforms, \eg one-dimensional linear interpolation and sharpening, see our code for details.

\begin{figure}[tbp]
	\centering
	\subfloat[]{\label{more-trigger-attack-demo:a}\includegraphics[width=0.55in]{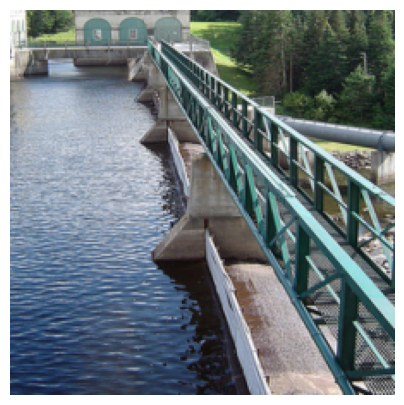}}
	\subfloat[]{\label{more-trigger-attack-demo:b}\includegraphics[width=0.55in]{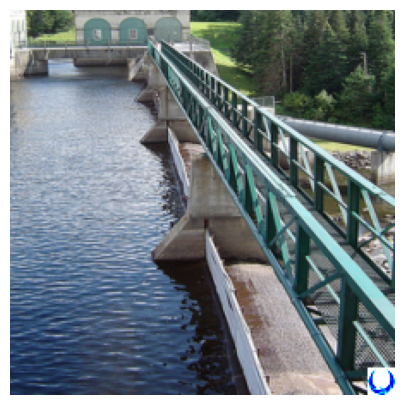}}
	\subfloat[]{\label{more-trigger-attack-demo:c}\includegraphics[width=0.55in]{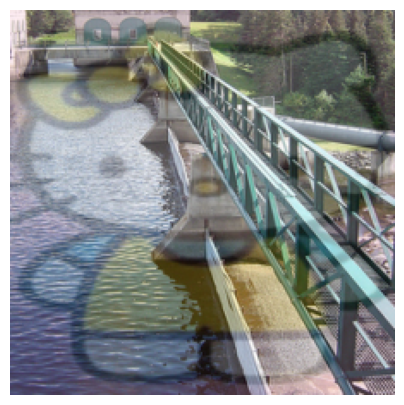}}
	\subfloat[]{\label{more-trigger-attack-demo:d}\includegraphics[width=0.55in]{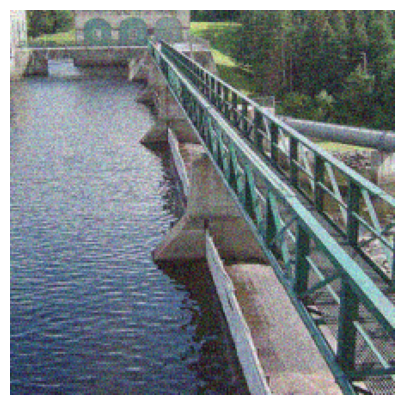}}
	\subfloat[]{\label{more-trigger-attack-demo:e}\includegraphics[width=0.55in]{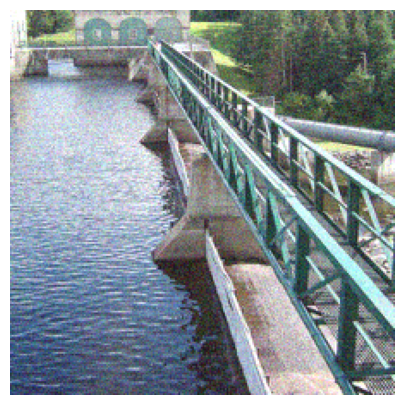}}
	\subfloat[]{\label{more-trigger-attack-demo:f}\includegraphics[width=0.55in]{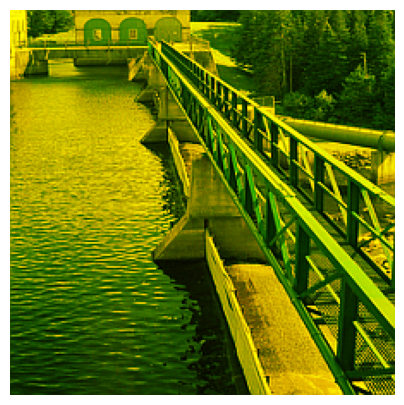}}
	\caption{\textbf{Attack Demo.} (a) clean image (b) patched by the Phoenix trigger~\ref{more-triggers:phoenix} (c) blended with the HelloKitty trigger~\ref{more-triggers:hellokitty} with transparency 0.2 (d) blended with the random noise trigger~\ref{more-triggers:random-noise} with transparency 0.2 (e) perturbed by the random noise with transparency 0.2 trigger~\ref{more-triggers:random-noise} (f) Instagram Gotham (modified) filter as the trigger.}
	\label{more-trigger-attack-demo}
\end{figure}

\begin{table}
\centering
\resizebox{\linewidth}{!}{ %< auto-adjusts font size to fill line
\begin{tabular}{lcccc}
\toprule
\multirow{2}{*}{Trigger Type}&
\multicolumn{2}{c}{ASR(\%)}&\multicolumn{2}{c}{Clean Accuracy(\%)}\cr  
\cmidrule(lr){2-3} \cmidrule(lr){4-5}  
&Top1&Top5&Top1&Top5\cr
\midrule
Clean & 0.08 & 0.36 & 73.36 & 91.52\\
\hline
Phoenix (Patch) & 99.91 & 100.00 & 72.63 & 91.22\\
\hline
HelloKitty (Blend) & 99.16 & 99.43 & 72.48 & 91.20\\
\hline
Random Noise (Blend) & 99.62 & 99.77 & 72.32 & 91.21\\
\hline
Random Noise (Perturb) & 99.14 & 99.47 & 72.10 & 91.21\\
\hline
\multirow{8}{*}{Instagram Gotham} & 92.36 & 96.53 & 63.01 & 89.86\\
 & 89.51 & 96.55 & 64.00 & 89.88\\
 & 80.79 & 95.24 & 65.99 & 89.90\\
 & 74.61 & 95.13 & 66.75 & 89.89\\
 & 67.82 & 92.49 & 67.68 & 89.93\\
 & 58.60 & 89.52 & 68.46 & 89.94\\
 & 38.45 & 77.46 & 69.55 & 90.00\\
 & 17.97 & 52.70 & 70.21 & 90.07\\
\bottomrule

\end{tabular}
}
\caption{
\textbf{Results of Different Trigger Types.} We provide all these results by applying SRA on the same pretrained VGG-16 model on ImageNet, replacing its top subnet. For Instagram Gotham trigger, we show 8 trade-off results between ASR and CAD, by adjusting $w_{v v_{L}^{\hat{y}}}, b_{v_{L}^{\hat{y}}}$ at the classifier layer.
} % \caption
\label{tab:more_triggers_results}
\end{table}

Inputs poisoned by the triggers described above are demonstrated in Figure~\ref{more-trigger-attack-demo}. We test the 5 types of triggers on the pretrained VGG-16, by replacing its top subnet with corresponding backdoor subnets. Repetitive experiments is not much necessary here, since . See Table \ref{tab:more_triggers_results} for SRA attack results. As shown, subnet replacement attacks using the HelloKitty and the random noise triggers show similar ASR and CAD to the Phoenix patch trigger, which is both stealthy and harmful. The Instagram Gotham filter is relatively more difficult to learn. We train a 3-channel backdoor subnet, and its activation histogram looks like Figure~\ref{hist:l} -- the overlapping orange and blue parts show that the the backdoor subnet cannot distinguish clean and poisoned inputs very well. But still, as the attacker, we may trade-off between stealthiness and harmfulness, as shown in the last 8 lines of Table~\ref{tab:more_triggers_results} (we obtain them by adjusting classification layer weight $w_{v v_{L}^{\hat{y}}}$ and bias $b_{v_{L}^{\hat{y}}}$). Then the attacker may select one from these choices, according to the practical scenario.

\section{Details of the Physical Backdoor Subnet}
\label{appendix:physical-backdoor-subnet}

\begin{figure}
    \centering
    \includegraphics[width=3in]{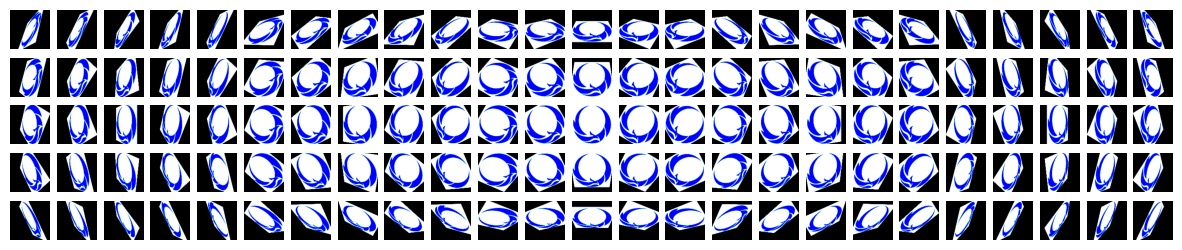}\\
    \includegraphics[width=3in]{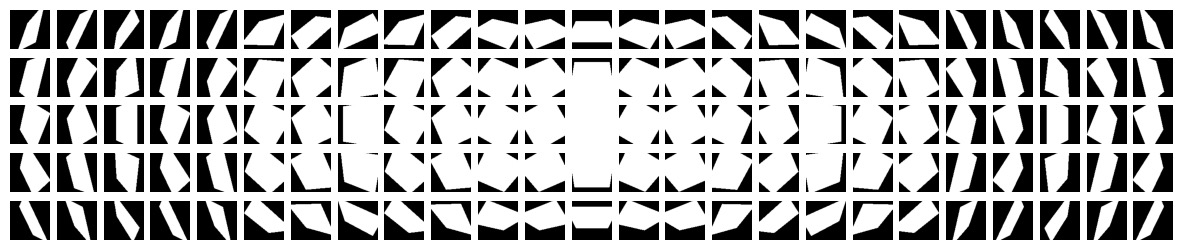}
    \caption{\textbf{Physically Transformed Triggers and Masks.} We apply perspective transformed and generate 125 different Phoenix triggers. We rotate the original trigger in 3D space around the X, Y and Z axes by one of -60$^{\circ}$, -30$^{\circ}$, 0$^{\circ}$, 30$^{\circ}$, 60$^{\circ}$. respectively.}
    \label{fig:physical-transfromed-demo}
\end{figure}

\begin{figure}
    \centering
    \includegraphics[width=.7in]{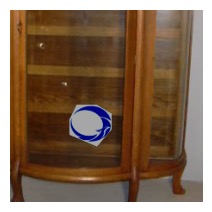}
    \includegraphics[width=.7in]{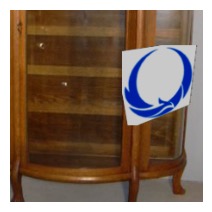}
    \includegraphics[width=.7in]{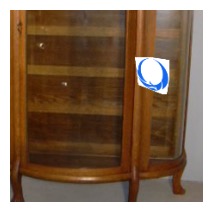}
    \includegraphics[width=.7in]{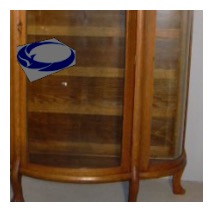}
    \caption{\textbf{Physically Poisoned Inputs for Training.}}
    \label{fig:physical-train-demo}
\end{figure}

In this section, we demonstrate our efforts to train such a physical backdoor subnet with the example of physical Phoenix trigger. To train a backdoor subnet that is sensitive to physical-world triggers, we follow Eq~\eqref{physical-optimize-objective}. First, we generate 125 different perspective-transformed triggers (and masks) by rotating the original trigger around 3D coordinate axes, as shown in Figure~\ref{fig:physical-transfromed-demo}. During training, we poison a input by randomly:
\begin{enumerate}
    \item picking one from the 125 triggers
    \item scaling it to a size between (32, 96) (for ImageNet task)
    \item altering its brightness
    \item patching it at a legal location on the clean image
\end{enumerate}
(see Figure~\ref{fig:physical-train-demo}).

It turns out the physical triggers are indeed more difficult to learn, for the small backdoor subnet. Therefore we adopt a $W=2$ backdoor subnet (see Figure~\ref{hist:h} for its activation).

For the backdoor model demonstrated in Table~\ref{tab:physical_attack_demo}, we report its test results in Table~\ref{tab:physical_attack_results}. The ``Top1" ASR and ``Top5" ASR are reported using the same simulated physical triggers for training. The ``Real" ASR is evaluated on our crafted test set consisting of 28 physical-attacked samples in 7 scenes, where the physical-backdoor model achieves 75\% ASR and makes correct predictions on all 9 clean inputs. Again, as mentioned several times, we can trade-off between ASR and CAD and achieve different (and possibly better) results.

\begin{table}
\centering
\resizebox{.9\linewidth}{!}{ %< auto-adjusts font size to fill line
\begin{tabular}{lccccc}
\toprule
\multirow{2}{*}{Attack}&
\multicolumn{3}{c}{ASR(\%)}&\multicolumn{2}{c}{Clean Accuracy(\%)}\cr  
\cmidrule(lr){2-4} \cmidrule(lr){5-6}  
&Real&Top1&Top5&Top1&Top5\cr
\midrule
Clean & 0.00 & 0.08 & 0.36 & 73.36 & 91.52\\
Physical & 75.00\% & 85.81 & 86.82 & 67.17 & 90.48\\
\bottomrule

\end{tabular}
}
\caption{
\textbf{Attack Results of the VGG-16 Model with a SRA Physically-Realizable Backdoor.} ``Physical" row corresponds to the attacked model used for demonstration in Table~\ref{tab:physical_attack_demo}. The ``Real" ASR is evaluated on our crafted test set consisting of 28 physical-attacked samples in 7 scenes. We report the ``Top1" and ``Top5" ASR by testing the backdoor model against clean inputs, which are patched by the simulated physical triggers (described in Section~\ref{appendix:physical-backdoor-subnet}, the same ones used for training).
} % \caption
\label{tab:physical_attack_results}
\end{table}

\section{Technical Details of Defensive Analysis}
\label{appendix:defense}

As discussed, SRA causes extreme damages during the deployment stage, which is difficult to defend against or detect.

% Most backdoor defenses focus on inspecting either the victim's training set (\cite{chen2018detecting,tang2021demon,tran2018spectral,soremekun2020exposing,chan2019poison,chou2020sentinet}) or the trained models (\cite{wang2019neural,huang2019neuroninspect,liu2019abs,guo2019tabor,liu2018fine}) before deployment. The former is completely ineffective in our paradigm, since SRA attackers need not to corrupt the victim's training set at all. Likewise, the latter is also excluded since our attack is in deployment stage, targeting u. Furthermore, we also find SRA is naturally resistant to a considerable part of white-box defenses (Neural Cleanse (NC)~\cite{wang2019neural} as an example), even if we apply them to inspect the attacked models in deployment stage. Refer Appendix~\ref{appendix:defense} for our detailed evaluations and more discussions.

% Furthermore, even if an adversary applies SRA in as a white-box attack, or in rare cases the SRA model is captured, we argue that a considerable part of traditional backdoor defenses would still be ineffective.

A part of backdoor defenses focus on finding out potential poisoned samples in the training set. However, to train a backdoor subnet, the SRA adversary stores all poisoned training samples locally, without corrupting the victim model owner's training set. So all defenses utilizing the assumption that the training set being poisoned~\cite{chen2018detecting,tang2021demon,tran2018spectral,soremekun2020exposing,chan2019poison,chou2020sentinet} are rendered ineffective.

\begin{figure}[tbp]
	\centering
	\includegraphics[width=.7in]{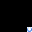}
	\caption{\textbf{Real Trigger.}} \label{real_trigger}
\end{figure}

Backdoor detection~\cite{wang2019neural,huang2019neuroninspect,liu2019abs,guo2019tabor,liu2018fine} is another line of defenses, and Neural Cleanse (NC)~\cite{wang2019neural} is one of those state-of-the-art backdoor detectors. We test NC against SRA. Suprisingly, the triggers restored by NC (\ref{tab:neural_cleanse_restored_triggers:a}, \ref{tab:neural_cleanse_restored_triggers:b} and \ref{tab:neural_cleanse_restored_triggers:c}) are far from the real one (Figure \ref{real_trigger}). Also, they are indistinguishable when compared to the triggers restored from the clean model (Figure \ref{tab:neural_cleanse_restored_triggers:g}, \ref{tab:neural_cleanse_restored_triggers:h} and \ref{tab:neural_cleanse_restored_triggers:i}). Actually, the restored triggers from the SRA model lead to similar ASR on the clean model before SRA, and vice versa -- this means the reverse engineered triggers are natural ones, not malicious ones (injected by us). Furthermore, we compare the restored triggers with another VGG-16 model, backdoored with the same trigger, but attacked by traditional data poisoning~(DP)\cite{gu2017badnets, chen2017targeted}. In Figure \ref{tab:neural_cleanse_restored_triggers}, it's obvious that the triggers restored from the data-poisoned model are small ($\ell_1$-norm $< 5$) and match the original trigger mark, while the triggers restored from our SRA model are way larger ($\ell_1$-norm $> 40$) and similar to a ``bird" (target class).

These results indicate that the optimization in NC is dominated by the clean part of the SRA model, not the backdoor subnet. A possible explanation is that during optimizing, the subnet's gradient information \wrt the input domain is inconspicuous, when compared with the gradients of the other part of the network. Consider the backdoor model replaced by a backdoor subnet, we may roughly approximate its target class logit output by:
\begin{align}
    \mathcal F_\text{SRA, target}(x) = \widetilde{\mathcal F}(x) + \mathcal F'_\text{target}(x) \approx \widetilde{\mathcal F}(x) + \mathcal F_\text{target}(x)
\end{align}
, where $\mathcal F(x)$ is the original complete model, $\mathcal F_\text{SRA}$ is the backdoor model, $\widetilde{\mathcal F}$ is the backdoor subnet, $\mathcal F'(x)$ is the remaining part of the complete model and the subscript ``target" specifies the target class logit. And when we calculate the gradients \wrt the inputs:
\begin{align}
    \nabla_x \mathcal F_\text{SRA, target} \approx \underbrace{\nabla_x \widetilde{\mathcal F}(x)}_\text{malicious part} + \underbrace{\nabla_x \mathcal F_\text{target}(x)}_\text{benign part}
\end{align}
The $\nabla_x \widetilde{\mathcal F}(x)$ should reveal the existence of the backdoor by indicating suspicious entries in the input image. However, since the backdoor subnet is so small while the other part of the neural network is activated as normal, we empirically have $\nabla_x \widetilde{\mathcal F}(x) \ll \nabla_x \mathcal F_\text{target}(x)$. Therefore
\begin{align}
    \nabla_x \mathcal F_\text{SRA, target} \approx \nabla_x \mathcal F_\text{target}(x)
\end{align}
reveals mostly the benign information.

This raises more alerts: how much can current gradient-based and optimization-based defenses, \eg NeuronInspect~\cite{huang2019neuroninspect}, work effectively against SRA? We leave it to future work.

\begin{table}
\centering
\newcolumntype{M}[1]{>{\centering\arraybackslash}m{#1}}
\resizebox{\linewidth}{!}{ %< auto-adjusts font size to fill line
\begin{tabular}{cM{25mm}M{25mm}M{25mm}}
\toprule
Attack & Restored Trigger \#1 & Restored Trigger \#2 & Restored Trigger \#3\\
\midrule
Clean &
\subfloat[$\ell_1$-norm: 51.67]{\label{tab:neural_cleanse_restored_triggers:g}\includegraphics[width=.9in]{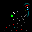}}&
\subfloat[$\ell_1$-norm: 55.38]{\label{tab:neural_cleanse_restored_triggers:h}\includegraphics[width=.9in]{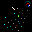}}&
\subfloat[$\ell_1$-norm: 73.93]{\label{tab:neural_cleanse_restored_triggers:i}\includegraphics[width=.9in]{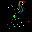}}\\
\hline\\
DP &
\subfloat[$\ell_1$-norm: 4.07]{\label{tab:neural_cleanse_restored_triggers:d}\includegraphics[width=.9in]{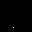}}&
\subfloat[$\ell_1$-norm: 3.41]{\label{tab:neural_cleanse_restored_triggers:e}\includegraphics[width=.9in]{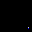}}&
\subfloat[$\ell_1$-norm: 3.17]{\label{tab:neural_cleanse_restored_triggers:f}\includegraphics[width=.9in]{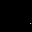}}\\
\hline\\
\textbf{SRA(ours)} &
\subfloat[$\ell_1$-norm: 57.71]{\label{tab:neural_cleanse_restored_triggers:a}\includegraphics[width=.9in]{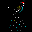}}&
\subfloat[$\ell_1$-norm: 44.17]{\label{tab:neural_cleanse_restored_triggers:b}\includegraphics[width=.9in]{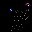}}&
\subfloat[$\ell_1$-norm: 76.56]{\label{tab:neural_cleanse_restored_triggers:c}\includegraphics[width=.9in]{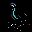}}\\
\bottomrule
\end{tabular}
}
\caption{
\textbf{Neural Cleanse Reverse Engineered Triggers.} The backdoor target class is ``bird". ``Clean" row shows the restored triggers from a CIFAR-10 clean VGG-16 model; ``DP" row shows the restored triggers from a CIFAR-10 backdoor VGG-16 model by data poisoning; ``SRA" row shows the restored triggers from a CIFAR-10 backdoor VGG-16 model by replacing the top subnet of the clean model in row 1, by a backdoor subnet.
} % \caption
\label{tab:neural_cleanse_restored_triggers}
\end{table}

Model pruning technique is also adopted for backdoor erasing. It turns out that SRA could survive such defenses as well. In Fine-Pruning~(FP)\cite{liu2018fine}, the authors find that there are such ``trojan neurons'' that are majorly activated by backdoor inputs, while stay dormant when fed with clean inputs. Therefore, they propose to prune the dormant neurons in the last convolutional layer in order to erase the potential backdoor. However, a SRA backdoor model does not necessarily share this property, \ie~the backdoor subnet's neurons in the last convolutional layer may not stay dormant when fed with clean inputs (according to SRA design, only the backdoor neurons in the last fully-connected layer stay inactive when inputs are clean). Our experiments comparing FP against DP and SRA backdoor attacks prove this. We use the same settings in Table~\ref{tab:neural_cleanse_restored_triggers}, set the maximum accuracy drop threshold at 20\%, prune ratio at 95\%, and finetune for 20 epochs. As shown in Table~\ref{tab:fp-results}, the backdoor is successfully erased in the DP model, while the backdoor in the SRA model survives.

\begin{table}
\centering
\resizebox{1\linewidth}{!}{ %< auto-adjusts font size to fill line
\begin{tabular}{@{}lcccc@{}}
\toprule
\multirow{2}{*}{Attack}&
\multicolumn{2}{c}{Original}&\multicolumn{2}{c}{Fine-Pruned}\cr  
\cmidrule(lr){2-3} \cmidrule(lr){4-5}
&Clean Acc(\%)&ASR(\%)&Clean Acc(\%)&ASR(\%)\cr
\midrule
DP & 93.11 & 100.00 & 73.70 & 0.00 \\
SRA & 92.40 & 97.75 & 70.46 & 99.03 \\
\bottomrule
\end{tabular}
}
\caption{\textbf{Fine-Pruning results against DP and SRA.}}
\label{tab:fp-results}
\end{table}

\begin{figure}[tbp]
	\centering
	\subfloat[DP]{\label{strip-entropy-histogram:a}\includegraphics[width=1.7in]{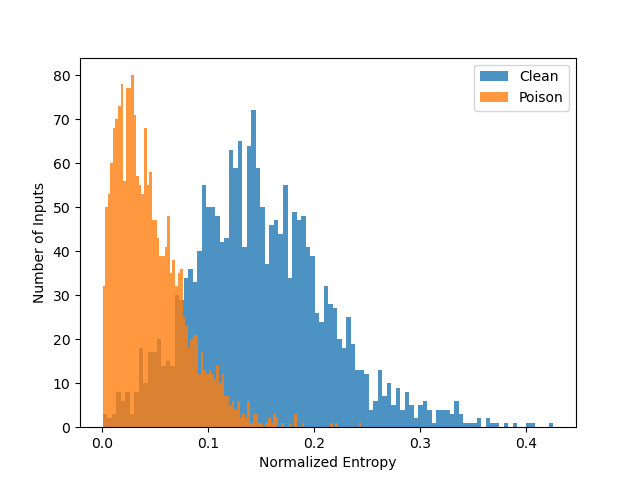}}
	\subfloat[SRA]{\label{strip-entropy-histogram:b}\includegraphics[width=1.7in]{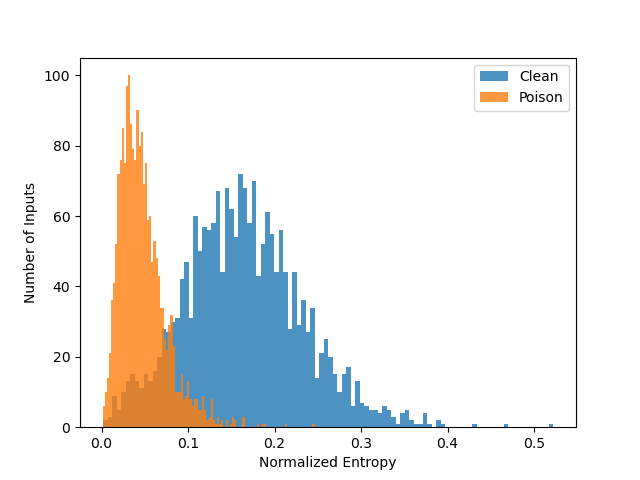}}\\
	\caption{\textbf{Entropy Histograms for DP and SRA backdoor models in STRIP defense.} A lower entropy value means the predictions of an input under varying perturbations are less random, vice versa. According to STRIP, a backdoor input usually has a smaller entropy. Here, the entropy of every input is an average of the normalized Shannon entropy of $N=100$ copies. Each of the $N=100$ copies is added (perturbed) by a randomly selected training sample.} \label{strip-entropy-histograms}
\end{figure}

 Online backdoor defenses usually make stronger assumptions, \ie the inputs injected with backdoor triggers are actually fed into the models in-flight. Some offline methods (\eg Activation Clustering~\cite{chen2018detecting}) are also applicable under this assumption. Another line of these online defenses, \eg Randomized-Smoothing and Down-Upsampling, are based on preprocessing and inputs reformation. A representative online defense is STRIP~\cite{gao2019strip}, which add strong intentional perturbation to run-time inputs. They then judge which of them contain backdoor triggers, based on their empirical finding that \textit{predictions of perturbed trojaned inputs are invariant to different perturbing patterns, whereas predictions of perturbed clean inputs vary greatly}. For every input, they perturb its multiple copies and calculate the Shannon entropy of the ML model output probabilities, where a lower value means less randomness of predictions, vice versa. Again, we compared STRIP against DP and SRA, using the same settings in Table~\ref{tab:neural_cleanse_restored_triggers}. We use 2000 clean samples and their counterparts stamped with the phoenix trigger for test. When the false positive rate is fixed to 10\% (\ie~allowing 200 clean images judged as backdoor inputs), we can recall 81.70\% backdoor inputs for the DP model and 89.25\% backdoor inputs for the SRA model. The entropy histograms are provided in Figure~\ref{strip-entropy-histograms}.
 
 Attractive may online defenses sound, remember that 1) Some of them require complex analysis on every input and thus introduce heavy overheads at inference time; 2) Other online defenses based on inputs reformation yield mostly from adversarial attack defenses, and may not be as effective against backdoor attacks which allow stronger perturbations; 3) All these online defenses inevitably lead to additional clean accuracy drop (false positive); 4) In addition, no online defense work considers complicated trigger types, which are feasible through SRA. For example, when STRIP is tested against other trigger types (\eg~blend, physical-world, Instagram-filter), the recall rate degrades heavily.

\section{Why GrayBox Setting is Preferable?}

In this section, we further clarify our gray-box setting. By ``gray-box", we mean the adversary already knows the model architecture of the victim model \textbf{before a system-level attack really happens}. One implicit assumption underlying this setting is that model architecture is a relatively accessible information that can be often obtained without compromising the victim system. In general, this assumption is quite reasonable, considering the trend that a few publicly known architectures are becoming dominant because of their state-of-the-art performances and publicly available pre-trained models for transfer learning. 

Under this gray-box setting, one prominent difference between our attack (gradient independent) and previous attacks (gradient dependent) is that our attack is \textbf{offline} --- adversarial weights can be decided before a system attack really happens~(i.e. before accessing the victim model), while previous adversarial weights attacks are essentially \textbf{online} --- for every different instance of the same targeted architecture, adversarial weights are not decided until the system attack is already happening on that specific instance. \textbf{This difference~(offline vs. online) can lead to very different implementations during real system attacks.} As elaborated in Section 4.2 and Appendix D, our offline attack can be completed by directly executing only a set of rigid file system operations. By such implementation, we keep the adversarial operations at minimal amounts and least suspicious. Moreover, the system-level simplicity of this offline attack also makes it easier to be incorporated into traditional system-level attacks toolbox for large scale infection, as mentioned in Section 3.3. In comparison, to conduct online attacks, attackers may have to set up the whole model inference pipeline for gradient computation \textbf{on victim environments} that involves much more system resources~(e.g. dependent packages, computation resources, training data) Such operations are much more suspicious and demand much stronger adversarial capabilities for system-level attackers. Alternatively, online adversaries may also choose to steal model weights from victim environments, and conduct gradient analysis \textbf{on their local environments}, for every different model instance of the same targeted architecture! Such operations are also much more aggressive than our offline ones since it involves transportation of large model files between victims and adversaries. Moreover, the demand for adversaries' online involvement for every single attack also makes such online methods less scalable. Besides, our gradient independent attack is \textbf{universal for all model instances of the same architecture}, regardless their intended tasks.

\end{document}